**Chapter Name:**

# Autonomous Vehicles: an overview on system, cyber security, risks, issues, and a way forward

**Author:**

(1) Md Aminul Islam, MSc, Advanced Computer Science, Post Graduate Research, Oxford Brookes University
(2) Sarah Alqhtani, BSc in EEE, Microsoft Intern, Field engineer in Honeywell, Post Graduate Research, Oxford Brookes University

## Table of contents

## 1. Introduction:

Throughout the years, autonomous vehicles (AV) have been in constant development. It is not a new technology, but since the 1990s the initiation for innovation started in the private and public sectors. These innovations include many systems for advanced driver assistance and some functional AV [1]. This technology creates a potential impact in travel behaviour, congestion, and safety. Other social impacts that this technology provides is fuel efficiency, parking benefits, travel time reduction, and crash savings, which is estimated to be 2000$ per year for every AV [2]. There are some challenges that may occur with AV like system performance in complex cluttered environments and safety while facing uncertain interactions with other participants in traffic. Questions regarding reliability and safety needs to be addressed, which can occur in the process of end-to-end learning and interactive planning. However, one of the most important purposes of the AV is to keep passengers and traffic safe. Most accidents happen due to human error, and on a yearly base, more than three thousand lives are lost daily [3].

In this chapter, the AV system will be discussed in detail. There are many sensors, technologies and algorithms that can be used to create different models of AV. However, within any technology some issues and risks will arise. Issues can vary from social, professional, environmental, legal and ethical, which will be discussed alongside a risk management assessment.

## 2. Autonomous Vehicles

Autonomous vehicles (AV) are usually not systematically organized. The vehicle system can differ in so many ways. Sensors are usually very different from one design to the other. Some designs completely depend on cameras, however other designs can also have a combination of technologies like laser scanners, milli wave radars, GPS receivers, and cameras. In autonomous vehicles, software issues should be addressed due to the large-scale platform. It is more efficient to build up the platform from scratch. The design process and application of algorithms such as path planning, vehicle control, and scene recognition needs a significant amount of integrated collaboration of knowledge and skills, which mostly demands engineering skills. Furthermore, the use of algorithms in autonomous vehicles demands huge databases to train the recognition process, and most importantly for localization maps of the public roads [4].

## 2.1. Sensors:

In every AV the main component that makes it autonomous is sensors. The fusion of data gathered from the sensors that should be accurately interpreted, along with the vehicle control system is the most important point in AV. For better visualization, the process of the AV system can be divided into four main categories (Figure.1). The first category is the sensors, which can be many different kinds that are installed in the vehicle. These hardware components are installed in the system to sense the surroundings of the vehicle. After that the sensors gather the data to translate and process them into meaningful information, which is the perception step. Furthermore, the output of the perception process is used to determine behavioral planning for the long-range and the short-range path plan. Then there is the control system that takes the path plan that is generated from the planning subsystem and sends to the vehicle control commands accordingly.

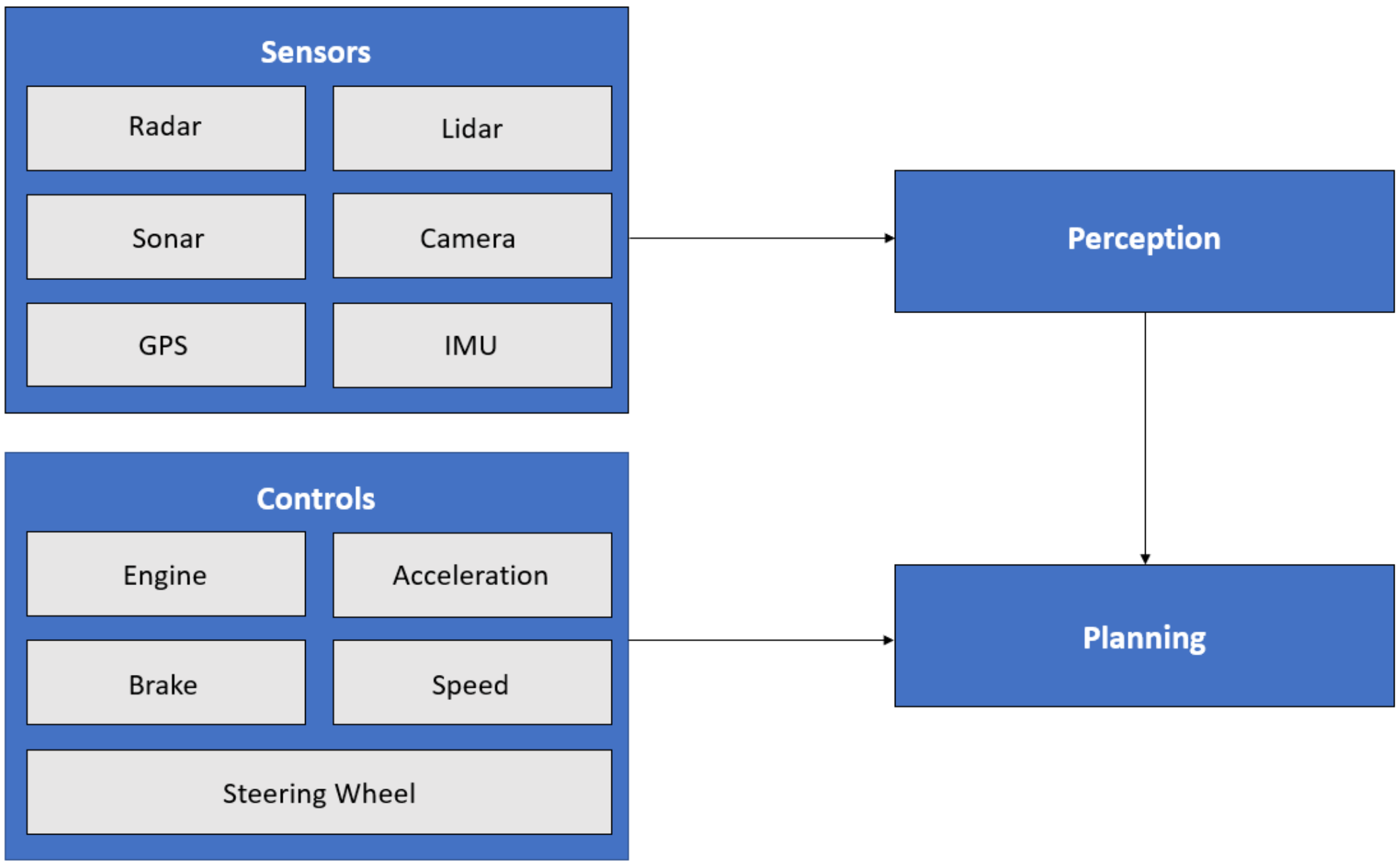

Figure.1 – Autonomous vehicle system Diagram

Sensors are a very important part of the AV, and different types of sensors are used in AV designs. The main types of sensors that are most commonly used in this system are radar, cameras, sonar, wheel odometry, lidar, inertial measurement unit (IMU), and a global positioning system (GPS). These sensors are used to collect data, process them in the impeded computing system and use them to control the speed, brakes, and the steer of the vehicle.

### 2.1.1. Camera:

Camera sensor is usually the preferred choice of sensors that are used in manufacturing an AV, and it is the first type of sensor to be used in this system. It is the best way to visualize the surroundings of the vehicle. There are many reasons why this type of sensor is commonly used, mostly because it is affordable, available, and has an accurate classification in interpreting texture. However, the use of the camera needs computational power to process the data. In the most recent high-definition cameras, the pixel per frame rate can reach up to a million, and this also includes 30 to 60 frames in a second. This means that to process in real-time, a multi-megabyte of data is going to be processed [5].

Unlike the other sensors, which are active, the camera or the optical sensor is a passive sensor. It is called a passive sensor because it gathers data in a way that is non-intrusive. Due to the affordability of a camera, vehicles can have it both in the rear end and in the front. This can enable a 360 degrees vision that can make tracking cars coming from a curve or switching lanes more efficient. Moreover, there are many applications that use visual information such as traffic sign recognition, lane detection, and object identification, which can vary to pedestrians, obstacles, and many other objects that can be found on the road. These applications do not cause any need for modification within the road infrastructure [6].

### 2.1.2. Radar:

Radars are defined as Radio Detection and Ranging. They are sensors that are embedded in the vehicle for different applications like collision warning and avoidance, adaptive cruise control, and blind spot warning [5]. These applications are achieved by using the radar in object detection, determining their position, and detecting the speed relatively to the vehicle. The progressed development in millimeter-wave semiconductor technology and signal processing techniques are a main factor in the radar system. Those techniques in signal processing made the resolution and estimation process better in every measurement dimension, including azimuth-elevation angles, velocity, and range of the surrounding objects around the vehicle.

The radar sensor has the ability to receive and transmit electromagnetic waves. The wave frequency band ranges from 3 MHz to 300 GHz. Like other sensors, the radar was made for information extraction that includes velocity, location, radar cross section (RCS), and range by using the reflected wave from the object targeted. The AV radar system operates at a frequency band 24 GHz and 77 GHz in the EM spectrum, which is called mm-wave frequency. These frequencies are used to achieve higher range resolution and velocity. The three main tasks of a fundamental radar operation are range estimation, direction estimation, and relative velocity.

**Range Estimation**

To get the range estimation of a target, the range R can be calculated based on the time delay of the round-trip of the EM waves. It is extracted from the target: $R = (c\tau/2)$, where the delay time of the round-trip is determined in seconds by the value $\tau$, and the value c is the speed of light in m/s ($c \approx 3 \times 108$ m/s). This $\tau$ value estimation will determine the range measurement. The radar transmits a type of EM waves that is necessary for the time delay estimation of the round-trip. One example is the pulse modulated continuous waves (CWs), which contains short and periodic power pulses and silent periods. The silent period enables the receiving process of the radar to get the reflected signals, and it also serves as a timing mark for the radar's performance of the range estimation. However, CW signals that are unmodulated such as $\cos(2\pi f_c t)$ lack those timing marks, therefore they are unused for range estimation. Furthermore, the reflected signal should arrive from the target before the start of the next signal. Hence, the radar's maximum detectable range relies on the pulse repetition interval $T_{PRF}$. Due to the loss and imperfection of the reflection path from the target, the signal that is transmitted and received from the radar undergoes attenuation. The radar electronics have internal noise that can affect the received target signal, alongside interferences that are caused by the reflected signal of other objects and human-made sources such as jamming. The ambient noise that is in the form of additive white Gaussian random process is the only thing that is considered in the problem of standard round-trip time delay estimation. The demodulation is assumed to already remove the carrier for the target signal x(t) at baseband to be modeled as:

$x(t)=\alpha s(t-\tau)+\omega(t),$

The value of α is a complex scalar that has a magnitude that represents attenuation that is caused by path loss, RCS of the target, and antenna gain. On the other hand, the ω(t) represents the additive white gaussian noise, which has variance $\sigma^2$ and zero mean. The goal here is to estimate the τ using the knowledge of the radar waveform that is transmitted s(t). With the assumption of finite energy $E_s$ and having a unit amplitude in the signal s(t), a radar receiver that is ideal can be using a matched filter with the impulse response $h(t)=s^*(-t)$. That will maximize the ratio of signal to noise ($SNR=(\alpha^2 E_s/\sigma^2)=(\alpha^2 T_p/\sigma^2)$) in the output. The correlation between the transmitted and the received signal pulses in this case is found by the matched receiver that is filter-based.

$y(\tau)=\int x(t)s^*(t-\tau)dt.$

The time delay maximum likelihood (ML) is estimated to be the time that the magnitude of the output's matched filter, which peaks at

$\hat{\tau}=\arg\max_\tau |y(\tau)|.$

The existence of the noise can perturb the peak's location, that will give an estimation error. The radar also needs to determine if the received signal has an echo signal from the target. Many strategies are being studied and developed when it comes to radar technology, but the most known decision strategy can be formed by statistical hypothesis testing, which leads to a simplified threshold testing in the matched filter output. Another main performance measure is range resolution, which distinguishes close spaced targets. If a nonoverlapping return signal in the time domain is produced, two targets can be separated in the range domain only. Therefore, the pulse width $T_p$ is proportional to the range resolution, which means that higher pulses give higher resolution. While less energy can be found in shorter pulses, which means it has a poor receiver detection performance and signal to noise ratio (SNR).

**Direction Estimation**

In the use of the wideband pulse like the FMCW, a discrimination of the target's velocity and distance is provided. The direction's discrimination is enabled by antenna array. In a case of traffic for example, there are usually several targets in the area surrounding the radar, which collects reflections that are direct and multipath. An angular target's location is estimated to deliver a comprehensive representation of the traffic scenery. Thus, in automotive radars, the description of the target's location is in the terms of a spherical coordination system $(R,\theta,\phi)$, where the $(\theta,\phi)$ coordination denotes respectively the azimuthal and elevation angles. If a single antenna radar is used in this case, in problems of range-velocity estimation, it might not be very sufficient because of the time delay measure $\tau=(2(R\pm vt)/c)$, which lacks the angular locations information of the target.

In order to make the direction estimation more feasible, the data of the reflected signals along multiple distinct dimensions should be collected by the radar. when a target is being located using electromagnetic waves in 2-D, the data of the reflected wave from the target should be collected in two distinct dimensions. Those two dimensions can be a combination of space, time, and frequency. For example, in a wideband waveform such as FMCW and in antenna arrays two unique dimensions will be formed. However, mm-wave bands, which is a smaller wavelength, correlate with smaller aperture sizes. Therefore, numerous amounts of antenna elements can be packed densely into an antenna array. Hence, a stronger and sharper beam, which is the effective radiation beam, in turn increases the angular measurements resolution.

**Velocity Estimation**

Based on the Doppler effect, the velocity of a target estimation can be found. Assume that a vehicle is moving forward in a differential velocity v. considering the relative motion between two vehicles, there is a delay in the reflected waves by time $\tau=(2(R\pm vt)/c)$. The received wave will experience a frequency shift known as the *Doppler shift* $fd=(\pm 2v/\lambda)$, which is caused by the time dependent delay term. Depending on the direction of the target, whether it is moving away or approaching, the sign is determined to be negative or positive. This sign along with the wavelength $\lambda$ are inversely proportional with the doppler shift. Although the shift in frequency is possible to detect using CW radar, it is not able to measure the range of the target.

In the case of FMCW radar, it transmits a periodic wideband FM pulses that has an angular frequency which increases linearly during the pulse. For the FM modulation constant K and the carrier frequency fc, the single FMCW pulse is shown as:

s(t)=ej2π(Jc+0.5Kt)t0≤t≤T.

The reflected signal from the target is combined with the transmitted signal to generate a beat signal with low frequency, which gives the range of the target. The process is repeated for P consecutive pulses. Waveforms that are two-dimensional represent a reflective pulse that are successive arranged in a two-time indices. The pulse number correspond with the slow time index p. however, the fast time index n is assuming that for every pulse, the continuous corresponding beat signal is sampled with the frequency fs to get samples N during the duration time T. if the single target is assumed and the duration of the reflected signal is neglected, the radar receiver FMCW output function of the two-time indices is shown as

d(n,p)≈exp{j2π[(2KRc+fd)nfs+fdpT0+2fcRc]}+ω(n,p).

A discrete Fourier transform through fast time n is applicable to get beat frequency fb=(2KR/c) coupled with the Doppler frequency fd. Another name for this operation is range gating or range transform. This operation is able to estimate the Doppler shift that corresponds with a unique gate range, which is possible by applying a second Fourier transform along the slow time. Using the two-dimensional Fast Fourier Transform (FFT) will enable finding the range-Doppler map [7].

### 2.1.3. Lidar

Light Detection and Ranging (LIDAR) sensors operate on infrared laser beams to estimate the distance between the sensor and an object. Lidar sensors that are currently used operate on light in the 900 nm wavelength range. However, longer wavelengths are used in some lidar sensors, which gives better performance in harsh conditions such as rain, snow and fog. The lidar uses laser beams in pulses across its field view, which are reflected by the objects within that view. It also has the ability to detect objects at a range that varies from a few meters to more than 200m. Yet, unlike the radar sensor, it may face difficulties in detecting objects that are at a very close distance. This sensor has a better spatial resolution than the radar sensor. That is because the lidar has a larger number of vertical direction scan layers, a denser lidar points per layer, and a more focused laser beam. It is able to collimate a laser beam with its short 905 to 1,550 nm wavelength, therefore it is able to make infrared (IR) spatial resolution on the order of 0.1 degrees. This will make extreme high resolution 3D characterization of an object within the scene of the sensor possible, without any significant back end processing. When it comes to detection in a low light condition, at night for example, lidar sensors can have a very high performance. Having said that, the lidar sensor lacks when it comes to directly measuring the velocity of an object and it must depend on various positions between two or more scans. It is also affected by the dirt on the sensor and the weather conditions in general [5][8].

There are many types of Lidar systems that can be applied to an AV. For the narrow pulsed ToF method, there are two types in Lidar beam steering systems. The first type is the mechanical Lidar, which uses a rotating assembly and high-grade optics to create a wide Field of View (FOV) usually 360-degrees. This aspect provides a high signal to noise ratio (SNR) within a wide FOV. Nevertheless, it generated a bulky implementation result. The second type is the solid-state Lidar. This lidar has a reduced FOV and no spinning mechanical components. Due to these features, this sensor is cheaper than other types. To make this sensor that has an FOV that can compete with the mechanical Lidar, multiple channels distributed in the front, rear and the sides should be applied to the vehicle, while fusing their data [8].

For the mechanical Lidar, also known as spinning lidar, the system has an IR-coherent light that is emitted from the laser. After that, the light is circularized and collimated to a round bean using optics. Every beam is put with a matching receiver, usually an avalanche photodiode (APD). Several pairs of the emitter detectors are mounted on a column that uses a motor to spin it, usually in frequency value between 10 and 20 Hz. The cycle's duty is low to guarantee eye safety. The number of pairs of emitter detectors stacked up vertically determines the vertical FOV. Whilst the speed of motor rotation and cycle duty determines the resolution of HFOV. This system is known to have the cleanest signals, a ratio of noise to date and gives a 360 degrees HFOV. However, the cost and size are a challenge in this system. Some doubts are also there about the need for self-calibration because the motor bearings wear [9].

There are several methods for solid-state lidar implementations, including:

**Microelectromechanical system (MEMs) Lidar:**

The MEMs Lidar system implements very small mirrors with a tilted angle that varies when inputting a stimulus like voltage. This system substitutes an equivalent of the mechanical scanning hardware, which is electromechanical. Typically, in this system the light collection receiver aperture, which determines the SNR received, is small making only a few millimeters. Cascading several mirrors is required in order for the laser beam to move in several dimensions. The alignment process is not significant, however, when installed it has a higher risk of vibrations and shocks, which is commonly experienced in a moving vehicle.

**Optical phase array (OPA):**

The OPA system is almost equivalent to the phased array radar. It contains an optical phase modulator that controls the speed of light going through the lens. This speed of light control process makes the optical wave-front shape control possible. A phenomenon occurs that would effectively steer the point of the laser beam to a different direction. It happens when the top beam does not experience any delay, while the other beams delay in an increasing amount. Other methods are also able to steer the backscattered light in the direction of the sensor, which eliminates the mechanical parts.

**Flash Lidar:**

The operation in the Flash Lidar is similar to the one in a standard digital camera, which applies an optical flash. A single pulse of large-area laser illuminates the surroundings in front of it and a photodetectors focal plane array placed close to the laser captures the back-scattered light proximity. The image location, distance, and reflected intensity is captured by a detector. Since in a single image, the entire scene is captured in this method unlike the mechanical laser scanning method, it has a faster data capturing rate because the entire image is captured in a single flash. In this method, the systema is more immune to the vibration effect that can tamper with the image. However, this method has a retroreflector that back-scatters a bit of the light and reflects most of it, which, in real life environment, makes an effect blinding the sensor entirely and makes it useless. Another downside is the existence of high peak laser power, which is necessary to detect from long distances and to illuminate the whole scene.

**Frequency-Modulated Continuous Wave (FMCW) Lidar:**

Unlike the other methods mentioned, this method does not use ToF principle, which is on using narrow light pulses. The FMCW Lidar adapts to the coherent method that produces a small chirp of frequency modulated laser light. The system is able to measure velocity and distance by measuring the frequency and phase of the returned chirps. The optics and the computational load are much simpler with the FMCW lidar, yet generating the chirps adds complexity [8].

## 2.2. Detection and Algorithms

A deeper understanding of the environment is needed for the AV to operate and react correctly. The roads in that environment are very hectic and unpredictable. The system should recognize objects and their location. The objects are usually vehicles, pedestrians, or bicycles. There are also the stationary objects on the road that the AV should recognize, such as traffic lights and signs. AV uses algorithms to classify its components. Some of those components are path planning, scene recognition, and vehicle control. Those classes contain a set of algorithms such as object detection, object tracking, and scene recognition that requires localization. The path planning for example, usually falls into motion and mission planning, however vehicle control falls into path following.

### Localization and Mapping

The AV needs a higher accuracy in navigating than the available GPS guidance systems, especially when navigating in a dynamic urban environment. The precision in localization determines the reliability of autonomous driving. An algorithm is used to enable this localization problem, which is the Normal Distributions Transform (NDT) algorithm, in particular the 3D module of it. The 3D NDT version applies a scan matching over 3D map data and 3D point cloud data. This results in a localization that can perform in an order of centimeter, giving high precision 3D map and high-quality 3D Lidar leverage. Not Only does the NDT algorithm have the ability to be used in 3D form, but it also has a computational cost that is not affected by the map size.

However, a simple way of applying localization is by obtaining data that helps determine the vehicle's location through observation of known and fixed beacons (points of reference) in an active environment. It is very straightforward in principle to get the position by giving a number of beacon observations. AV uses a wide range of systems that are beacon based for the localization problem. In specific, the factory AV use of the laser system and reflective beacons is very common. The beacons are based on ultrasonic transducers, radar reflectors, and microwave tags [10][11].

Another algorithm that is used to help with the localization problem is the Kalman Filter algorithm (KF). This algorithm is proven to be able to deal with complex localization problems in a very simple way, yet it has many limitations. In the KF algorithm an estimate of the point of interest is made by a recursive linear estimator, which utilizes a model of the observation of the process and a model of the process under consideration. It is usual to start the localization problem with a discrete time process model, which describes the platform's motion as:

$x(k) = f(x(k - 1), u(k)) + w(k)$

The state or location of the vehicle at a time k is described as x(k), and the average input or drive signals of the vehicle at a time k is described as u(k). The f is the kinematic function, which describes the vehicle's transition in the time between k-1 and k. whilst the w(k) presents the value of a random vector that describes the motion uncertainty of the vehicle in time. Observation of the vehicles location at a certain time is made, typically by getting the location of the beacons measurements that is applied in known places around the environment, according to the observation equation:

$z(k) = h(x\{k)) + v(k)$

The z(k) presents the observation that was made at time k, and the h presents a function relating the current location of the vehicle to the observation. The v(k) on the other hands is described as a random vector of the error that occurs in the observation process. The KF computes an estimated vehicle state recursively through a combination, that is linear, of the prediction based on the process model:

$x(k \text{ I } k - 1) = f(x(k - 1 \text{ I } k - 1), u(k))$

however, with the correlation that is based on the difference of prediction and true observation is shown as:

$x(k \text{ I } k) = x(k \text{ I } k - l)+W(k) [z(k) - h(x(k \text{ I } k - 1)]$

The W(k) is the gain given by the observation and the relative confidence in the prediction of the vehicle.

There are many features for using the KF algorithm, including a continuous confidence in the estimate of the location through the matrix of state of covariance, the ability to estimate a none directly observed state, the ability to describe in a single coherent form a variety of sensors, and the incorporation, incremental, and recursive of information into the estimate of a vehicle's location. Furthermore, the KF algorithm also has some limitations that should be considered. The first limitation is the need of accurate models for both observations and vehicles. Also, the necessity of linearization assumptions having a linear predictor corrector form of the updated equation. There are also the terms of the true noise that should be bounded by the first two moments of the assumed noise model and it should be well behaved. The last limitation is the form of the model applied, which is an analytic form, makes the capturing process of many physical environments and sensor properties difficult [12].

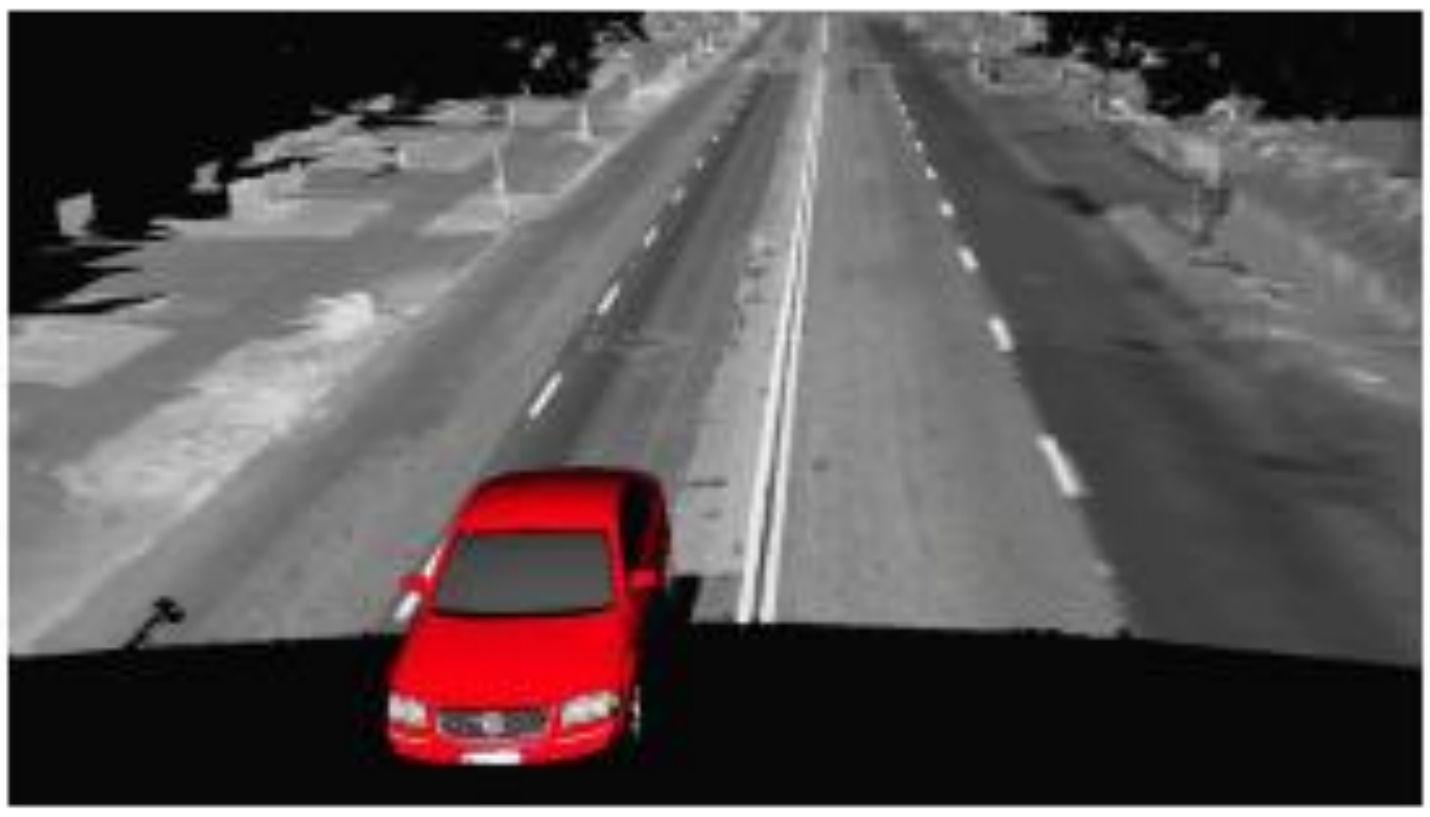

Figure.2 - Visualization of the scanning process. The Lidar scanner acquires infrared ground reflectivity and range data. Thus, the resulting map is 3D infrared images of the ground reflectivity [13].

## Object Detection

The object detection algorithm is used in AV to avoid accidents and abide by the traffic rules. In specific it is used to recognize vehicles, pedestrians, bicycles, traffic lights, and traffic signs. There are man features that are desired in an object detection system, such as accuracy, real-time capable, able to learn without the need for massive hand-labeled training sets, inherently multi-descriptor, inherently multi-class, fewer manual feature engineering required, and can also add new object classes and descriptors with no relearning from scratch.

To subtract deeper information, the images get segmented. This results in the removal of the local ground plane and the components connected in clusters of the remaining points , which is applied on a 2-D grid for efficiency. the segment is inputted into a standard tracker that includes velocity and position in its state variable. This track classification is important for high performance. Two boosting classifiers are applied to achieve track classification. One is the motion descriptors of the whole track, and the other is the object's shape in each frame of the track. Bayes filter is used to combine these predictions, and it is shown as the frame descriptor in Figure.3 for three objects, a pedestrian, a car, and a bicyclist [11][14].

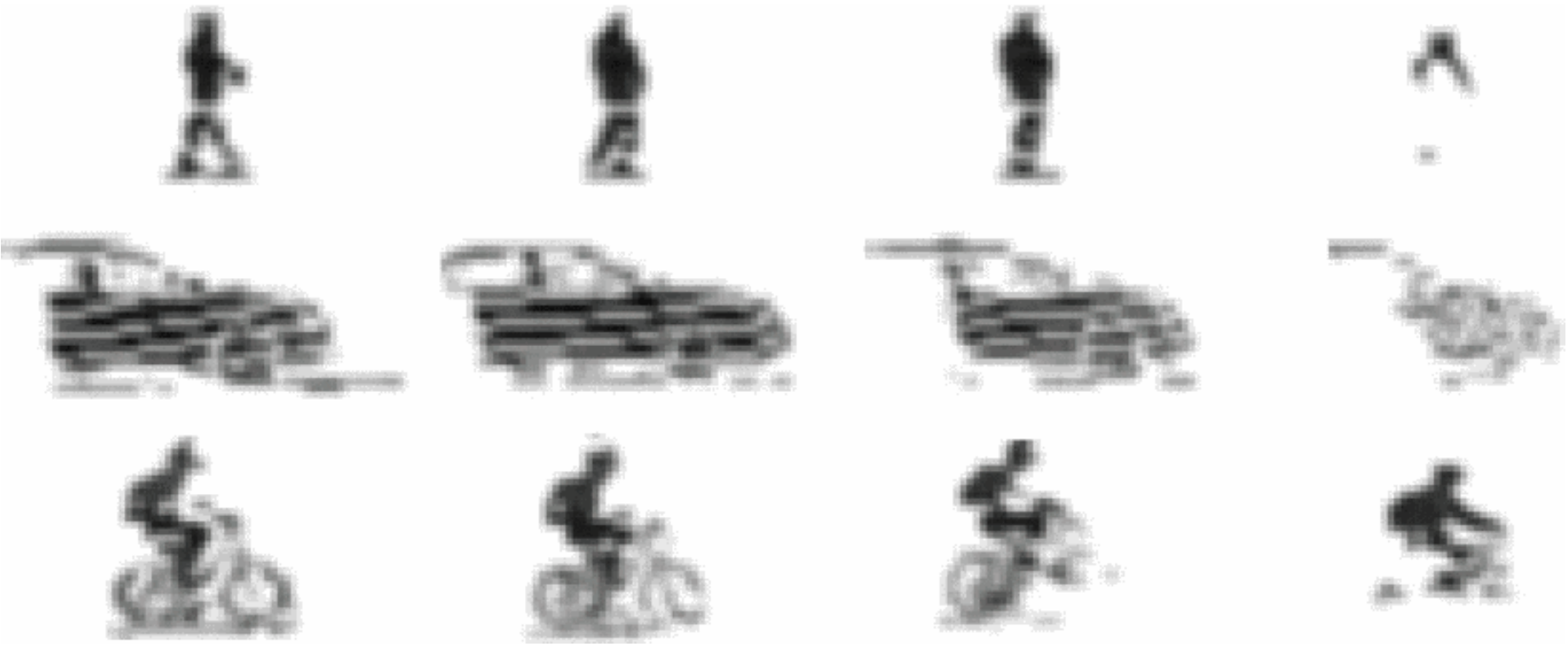

Figure.3 - recognized virtual camera intensity images for three objects tracked. Segmentation depth of the object provides invariance to the background clutter. The depth classification data enables objects addition in a canonical orientation, therefore it provides some perspective invariant measures [11].

When it comes to object recognition, there are two methods, the supervised method and the semi-supervised method .

**Supervised method**

The system is real-time, accurate if given a good tracking and segmentation, and inherently multi-descriptor and multi-class. Laser-based objection detection algorithm on junior has three components, which are segmentation, tracking, and track classification. The first step, the object gets segmented from the surrounding environment using depth information. After that it gets tracked with a KF. Therefore, the tracking and segmentation methods are model free, which means that no object class models are applied during the stages. The tracked object's classification is achieved when applying a boosting algorithm across a few high dimensional descriptors spaces, which encode shapes, motion properties, and sizes [15].

The largest error source in the system can be found in the segmentation and tracking process. Objects like bicycles and cars continuously avoid being segmented together with the surrounding environment, yet when it comes to pedestrians and other objects of interest this is not the case. When an object gets closer to another object the system generates a false negative.

In the meantime, to maintain a real-time capability, the system needs to remain feed-forward, because more mathematically complex methods that simultaneously take segmentation, tracking, and classification into consideration do not have real-time capabilities. When only classifying pre-segmented objects, the system takes more time on each processing each candidate than the sliding window system prevalent in computer vision for example.

**Semi-Supervised Method**

Segmentation and tracking that are model free allows a highly effective learning object model methods disregarding the need for huge numbers of hand-labeled data. This is called tracking based on a semi-supervised method. This method actively learns a classifier and collects new and useful training instances by applying tracking information. In bicycles, for example, the method can learn to identify half occluded bicyclists from tracks that are unlabeled, which includes views that are half-occluded and occluded. The method had been tested to show an accuracy that is relatively high of track classification with three hand labeled training tracks for every object class only [14].

## 2.3. Cloud Robotics

The term cloud robotics has been introduced by James Kuffiner from Google in 2010. This term illustrates a new method that allows robotics to use the internet as a source for large parallel computation and exchanging massive data resources in real time [16]. It enables access on-demand to almost an unlimited computational resource, which can come in handy when dealing with bursty computational workloads that need periodically a large amount of computation. The idea of a remote computer in robotics in general is not new, but it gave so many different possibilities for applications in mobile robot systems like the AV. The AV system can access large-scale map data and images through the cloud, which eliminates the need for local data storage. It also gives the ability for the AV systems to communicate with each other. However, some challenges would come up when using remote cloud resources, particularly when using commodity cloud facilities like Microsoft Azure and Amazon web services. Those services may bring some variables that are beyond the robotic system's control. There is also the issue with communicating with remote clouds, which can cause unanticipated delay in the network. The time of cloud computation can also depend on available compute resources and the number of jobs that are running in that system at this time. In that case, even though the cloud can achieve real time performance in a normal case, in the case of times that it may get overloaded, latencies may occur and affect the onboard processing which is needed for critical tasks. However, target recognition can be moved to the cloud, while maintaining the stability control, short-term navigation, and low-level detection local. Using this hybrid method can lower the cost to detect many objects in near real time, while limiting the negative consequences when the target can not be met in real time [17][18].

This method is applied in some vehicles like the Google autonomous driving project. The system is enabled to access images and maps that are collected and then updated into the satellite, crowdsourcing, and streetview from the network, which would generate accurate localization. Another example for this application is the Kiva system, which uses a new approach to warehouse logistics and automation by applying a huge number of mobile platforms in order to move pallets to update tracking data and coordinate planforms by using a local network [16].

## 2.4. Internet of Things (IoT)

An integral part of IoT is large-scale application of interconnected actuators and sensors. There are many applications made possible with IoT technology, among them is smart transportation systems. Those systems that apply communication, control technologies and computing together are called cyber-physical systems (CPS). This system requires real time processing, because the sensor network is integrated with the control of the physical system. This technology provides a system where vehicles can communicate. This can be used in intersection management applications. However, this application requires secure, low-latency, reliable, and a short-range communication [19].

After the invention of IoT an emerging concept came called the internet of vehicles IoV. This was the result of the advancements made in AI, wireless networks, and sensor technology. This technology already established architectural proposals among which is Cisco. The Cisco architecture includes four layers. The first is the end point, which contains the vehicle's software and hardware. The second layer is infrastructure, which defines all technology that enables connections. The third layer is operation, which monitors the flow-based management and the policy enforcements. The last layer is the virtual layer, which contains all different types of cloud subscriptions [20].

## 2.5. Blockchain

In a scenario where AVs are part of an accident, how do these accidents get recorded to determine the cause and liability? When it comes to recordings, it is also important for this data to be trusted, verified, and not tampered with. To achieve these properties blockchain technology can be used for event recording schemes, which can be useful for the forensics system. The blockchain contains a series of blocks, each block has a set of transactions that are timestamped and a hash of the previous block [21]. Originally blockchain technology is made for bitcoin, which is a digital cryptocurrency.

Recording the accident as a timestamp transaction would be saved in real time into a new block. AV has the capability to save the vehicle from accidents in real time, however it is not always possible due to complex solving of the hash puzzle [22]. A hash function takes a set of inputs of different arbitrary sizes, then puts them into a data structure like tables that has elements with fixed sizes [23]. One of the recording mechanisms to save events into a block is the Proof of Event with Dynamic Federation Consensus. In an accident, the directly involved vehicles broadcast an event generation, which can be via IEEE 802.11p [DSRC] for example, and only the vehicles that have DSRC communication range can respond and receive [22]. DSRC (Dedicated Short Range Communication) is 802.11p based wireless communication technology that provides direct communication between surrounding infrastructure and vehicles with high speed and high security without interrupting surrounding cellular infrastructure [24]. After that, both involved vehicles will receive a command to generate and broadcast the event to the vehicular network that is defined by the existing cellular network infrastructure. A random federation group in the vehicular network is created to save and clarify the accident's data into a new block (Figure.). This is done by using a multi-signature scheme. The last steps are saving and sending this block to the Department of Motor Vehicles (DMV) to be saved in the permanent record. This mechanism uses the accident's data collected from different sources and the generated hash digest to protect data trustworthiness and integrity. The recordings also provide traceable evidence [22].

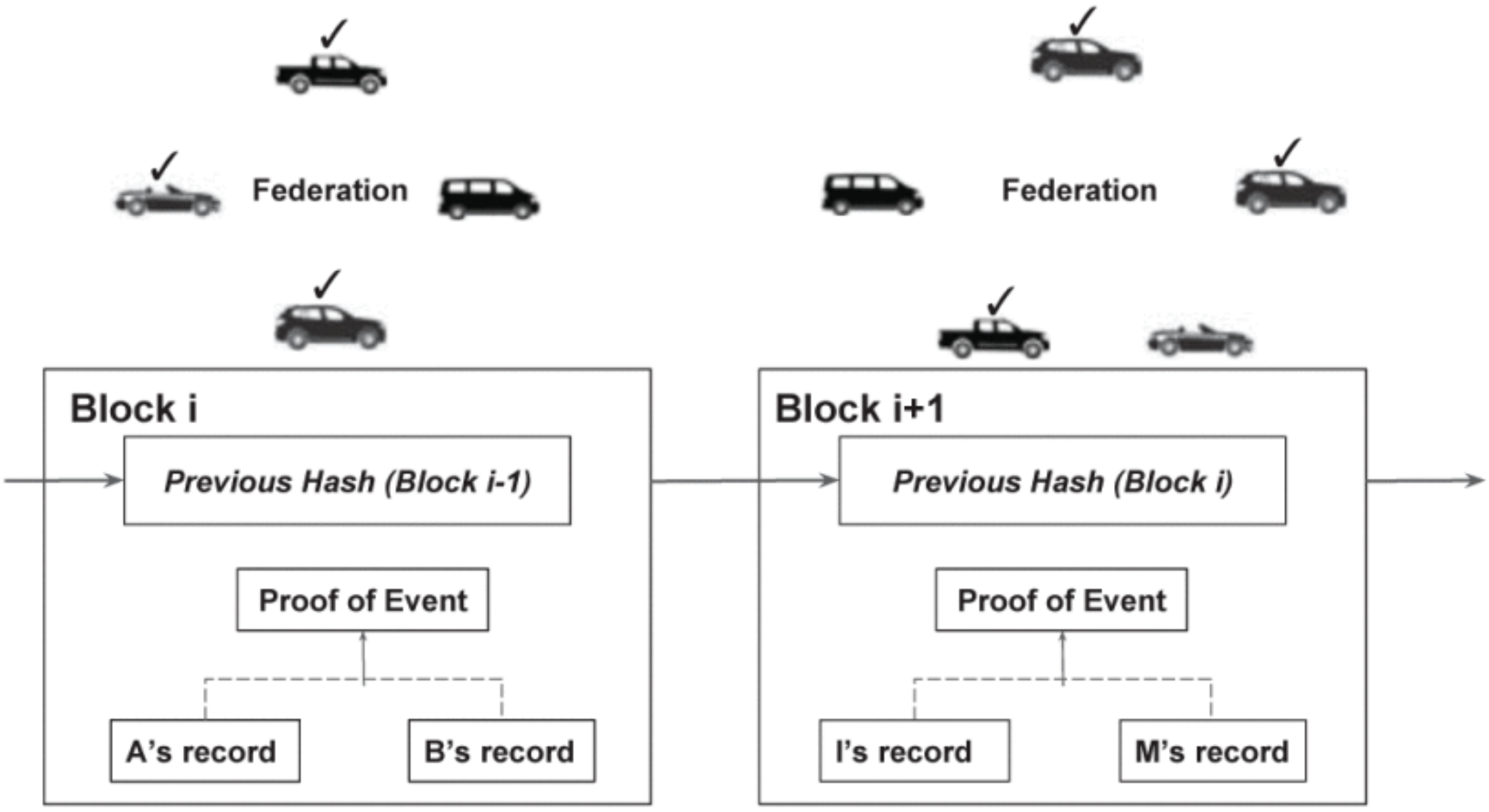

Figure.4 - The blocks contains hash values and data from the previous block, and it is all confirmed by verified federation vehicles [22].

## 3. Existing Models of Autonomous vehicles

The myth or science fiction of self-driving cars is now a reality as a fastest growing market where riders will enjoy hands free feeling in a safer way. The market will rise from its current status of US$54.23 billion to US$556.67 by 2026 and 33 million cars will be on road by 2040. [31], [32]

Some brands are running their operation and some are going to start like Tesla, Nuro, Zoox, Baidu, Nissan, GM, Aptiv etc. which have been shown in the table below: [33]

Table 1: Some AV Companies

| Name | Cruise | Waymo | Voyage | Nodar | Seeva | ABB | Reality AI |
|---|---|---|---|---|---|---|---|
| Founded Year | 2013 | 2009 | 2017 | 2018 | 2016 | 1988 | 2015 |
| Home Base | San Fransisco | San Fransisco | San Fransisco | Boston | Seattle | Colorado | New York |

## 4. Traffic Flow prediction in Autonomous vehicles

With the increase of the living standard, a higher need for the transportation industry occurred. There is a higher number of transportation users due to the growing population. This resulted in an increasing traffic flow and congestion, which is causing a huge problem. In order to solve the traffic problem effectively, the intelligent transportation system (ITS) technology can be applied. This technology is a transportation management system that merges different technologies like artificial intelligence, image analysis, global positioning, electronic information, and many other technologies. Furthermore, ITS technology is considered to be an effective way to solve transportation problems. It also provides a unique performance, which reduces traffic pollution and accidents, and solves traffic congestion in a non-traditional way that is not owned by any traditional method. Two roles that are of great importance in the ITS is traffic control and guidance [25]. There are also two main theories to guarantee a traffic flow guidance system with a good operation. The first one is accurate traffic prediction of the data stream, and the second theory is traffic prediction techniques. The traffic flow is known to be time-varying, random, and nonlinear, which makes it hard to predict long-term traffic flow.

In the urban areas the traffic flow system has characteristics that are chaotic. Taking this idea into consideration the approximate restoration of the original system prediction is established by a nonlinear mapping construction, which establishes a predictive model. There are various researchers that have been conducted to establish this model, including methods like K-nearest neighbor (KNN), Bayesian network, radial basis function (RBF), neuro-fuzzy system, type-2 fuzzy logic approach, binary neural network, Kalman state space filtering, autoregressive integrated moving average, and many other models. The Neural Network in these prediction models is the main focus for a lot of experts. This is because neural networks have characteristics that are unique such as distributed storage and massive parallel structure. It also has a good self-adaptability, self-organization, incorrectness tolerance, ability to classify patterns, and a powerful function approximation. Yet, the performance of the network can be affected by improper preferences. The selection of width value and center vectors of the neural network's hidden nodes impacts the generalization and learning abilities greatly [26].

Another algorithm that can be used is the Artificial bee colony (ABC). The algorithm can be used for parameter's combination optimization of the neural network, due to its smaller control parameter and simple structure, easy implementation, and high coverage speed. However, in that case the algorithm is prone to premature coverage and slow coverage speed.

The ABC algorithm is inspired by the behaviors of honeybees, and the algorithm is assumed to be a colony that has groups of onlookers, employed bees, and scouts. Employed artificial bees make the first half of the colony, and the onlookers make the other half. There is one employed bee for every food source, which means that the number of food sources should be equal to the number of employed bees. The scout, on the other hand, is a former employed bee that left the food source. In general, the employed bee determines the food source from the neighborhood memory, then they forward this information to the onlookers which can also have one food source. The onlookers have food sources from their chosen neighborhood. After the employed bee leaves the food source and becomes a scout, it starts looking for a new food source. The algorithm can be summed up in four main steps:

1) determine the nectar amounts of the employed bees and move them to their food source.

2) Determine the nectar amounts of the Onlooker and move them to their food source.
3) Move scouts so they can search for new food sources.
4) Store the best food sources into the memory.

The possible problem solutions to be optimized are represented by the food source. The quality of the solution is represented by the nectar amount. A method called roulette wheel selection is applied to put onlookers on their foods. On the other hand, the scouts do not have any guidance when looking for food. They just need to find any kind of food source. This behavior causes low average food sources and low search cost. Scouts sometimes stumble on unknown and rich food sources. The classification of scout bees is controlled by limit control parameters. The limit parameter is equal to the number of trials for releasing a food source, which is a significant control parameter of the ABC algorithm [27][28].

## 6.0 Cybersecurity Risks (ISO 27005):

The AV must work in an integrated network system where it will be connected to internet through GPS, cloud, GPRS, or satellite. However the networking system must contain traditional infrastructure like data center, servers, databases, and so on. It is also provisioned with IP addressing and subnet system. The whole system will contains several types data like travel history, maps, riders data, owners profile etc. So it must comply information security where information risks will arise. To resolve this we can go through Information Risk Management (ISRM). The subsequent topic will focus on network design and ISO 27005.

## 6.1 Network Design

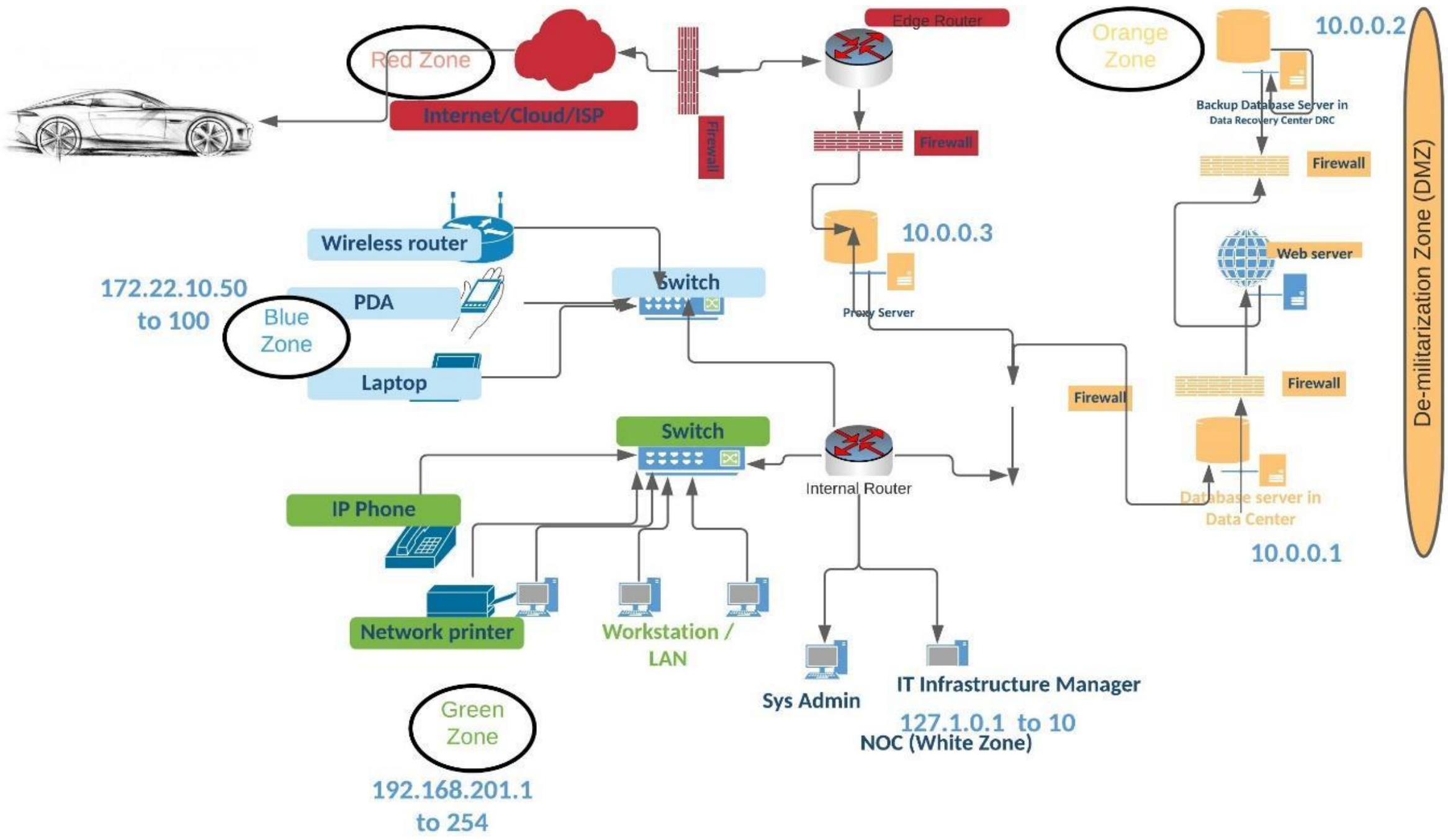

Figure 5: Model Network Design with IP Address and Zone (by Lucid Chart)

From the above diagram, it can illustrated that the entire autonomous vehicle system must work according to a secure network architecture. The whole network should be in an enterprise where no compromise for security ever permitted. Both of the Network Operation Centre (NOC) and Security Operation Centre (SOC) would be coordinated from a central hub. All possible aspects of zoning, proxy server, IP addressing, router, switch, firewall security system where all modern encryption, SSL, TLS, IMAP, SNMP, POP, secure protocols would be incorporated as the whole system deals with life of human while carrying them in road. The security is a prime concern thereby.

### 6.2 Risk Management- ISO 27005

Almost all companies who consider their growth with positive branding, sustainability and customer orientation, they always follow the security as first of everything where it matters. That's why all organization are coming forward to abide by the international standard guidelines. The ISO/IEC is the most acceptable standard on information security specially the ISO 27005 deals with Information risks. [34]

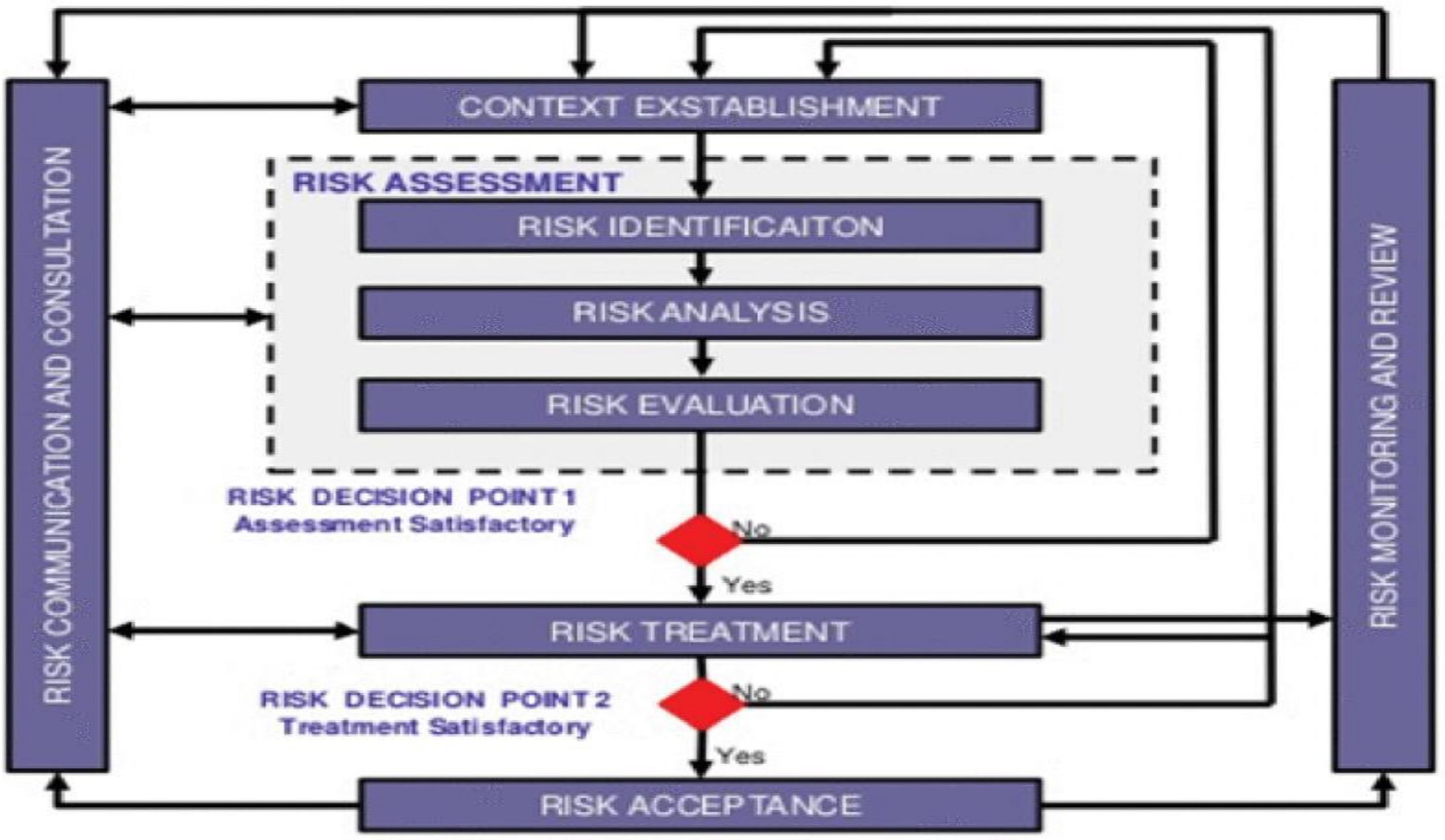

Figure 6: Information Security Risk Management Process [35]

In our above stated system we will go forward by following the above diagram in Fig 3.

This diagram will be discussed in detail in the subsequent topic having both in general and some focus on AV.

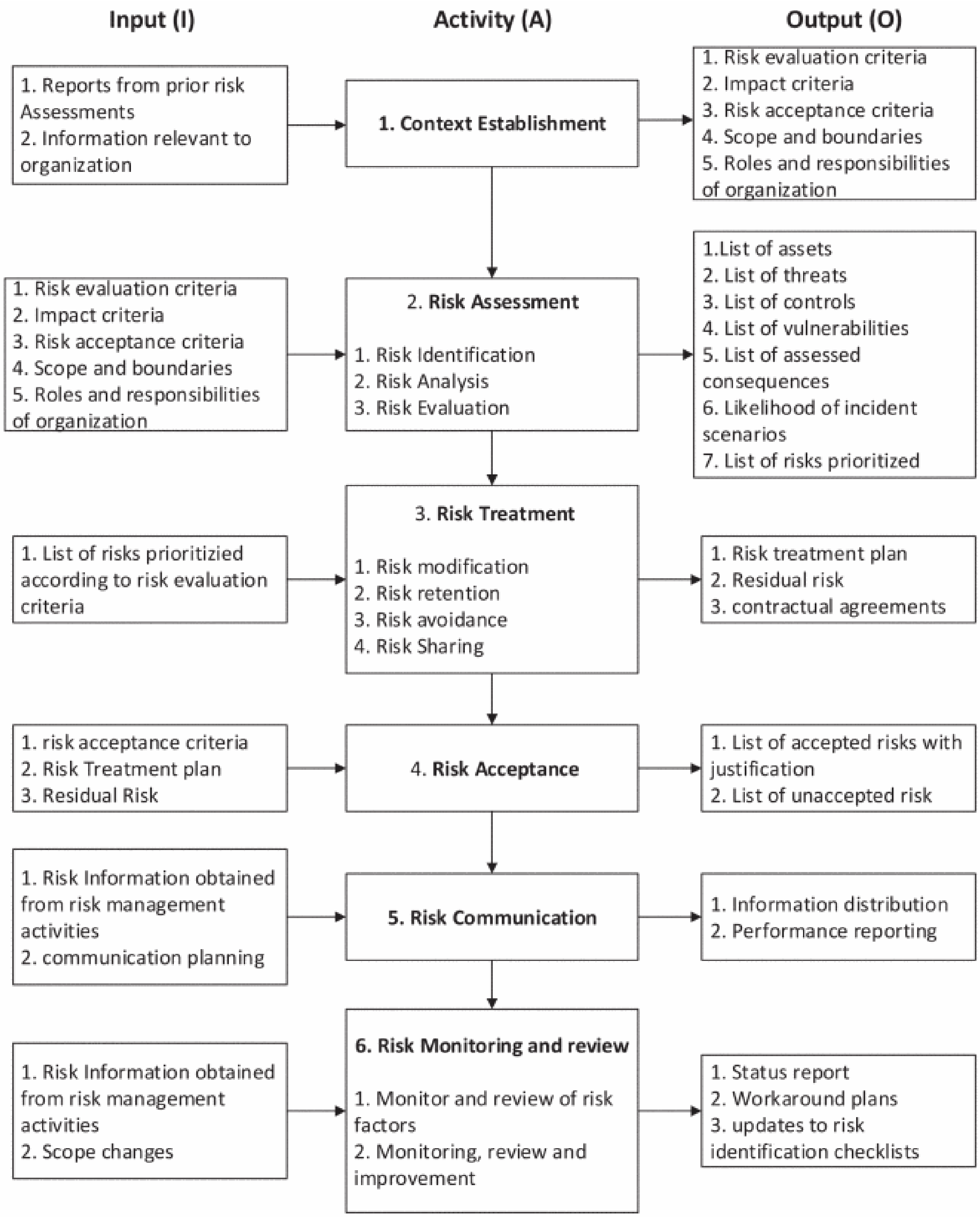

Figure 7: ISO27005 standard based on input, output of each activity [36], [37]

**Some Terms of IT Security Risks**

There are several terms which are very relevant to both technical and non-technical staff of the organization in order to implement ISRM properly which are discussed shortly below. [38]

- **Threat**: Any probable danger which may happen in future.
- **Vulnerability**: Any weakness what can be exploited by any means.
- **Breach**: If threat becomes successful through vulnerabilities.
- **Risk**: When loss is caused as low, high, or medium by any breach.
- **Relative risk**: It signifies multiple risks related to each other.
- **Temporal risk**: This type of risks can be protected by adequate measures.
- **Residual risk**: It means those which are still present after measures.
- **Compromise**: It happens when vulnerabilities are not considered.
- **Mitigation**: It is tend to repairing activities against of different risk elements.
- **Countermeasure**: It refers to those threats which are immitigable slightly different from mitigation. [39]

**Risk Calculation Models**

There are several models to calculate risk from different aspects which are discussed below:

• **CRAMM** (CTCA risk analysis and management method) is the risk analysis method which involves assets, threats, vulnerabilities associated with risks, countermeasures, implementation and audit.

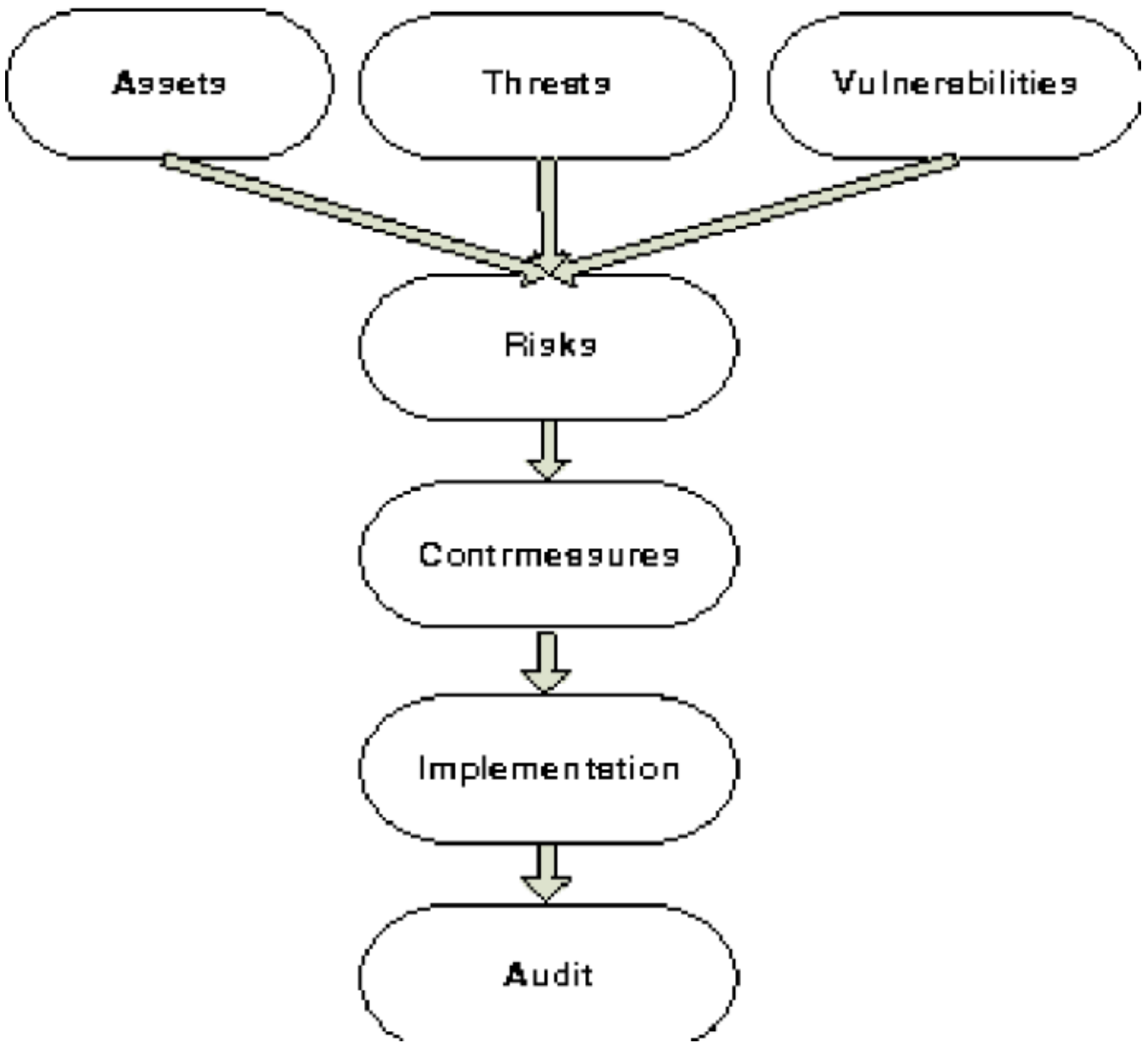

Figure 8: CRAMM risk assessment method [40]

• **DREAD**: Microsoft proposes the following straightforward DREAD model to practice:

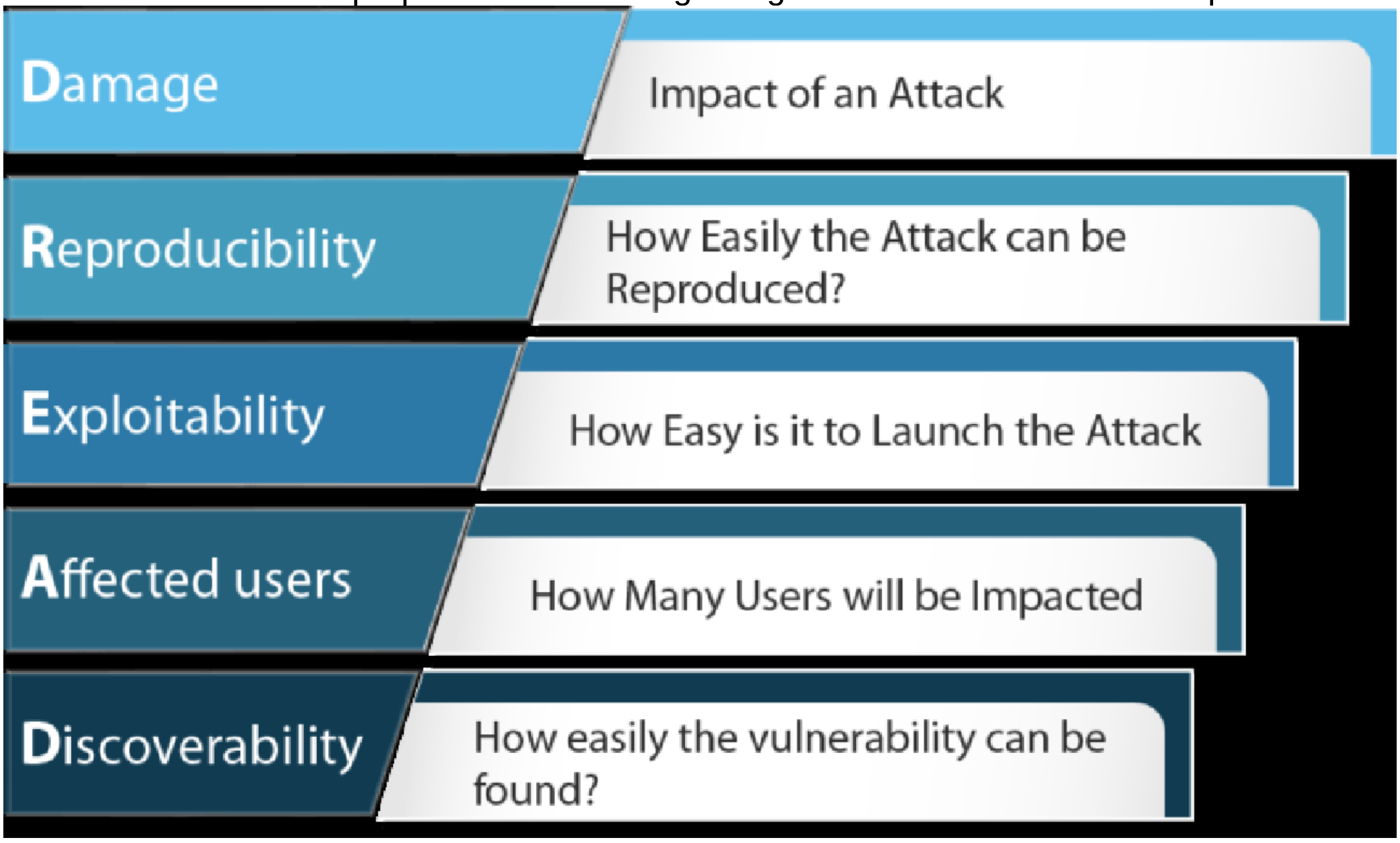

Figure 9: Dread Model, threat modeling, EC council [41]

• **STRIDE**: This model is used along with DREAD model for live systems in internet.

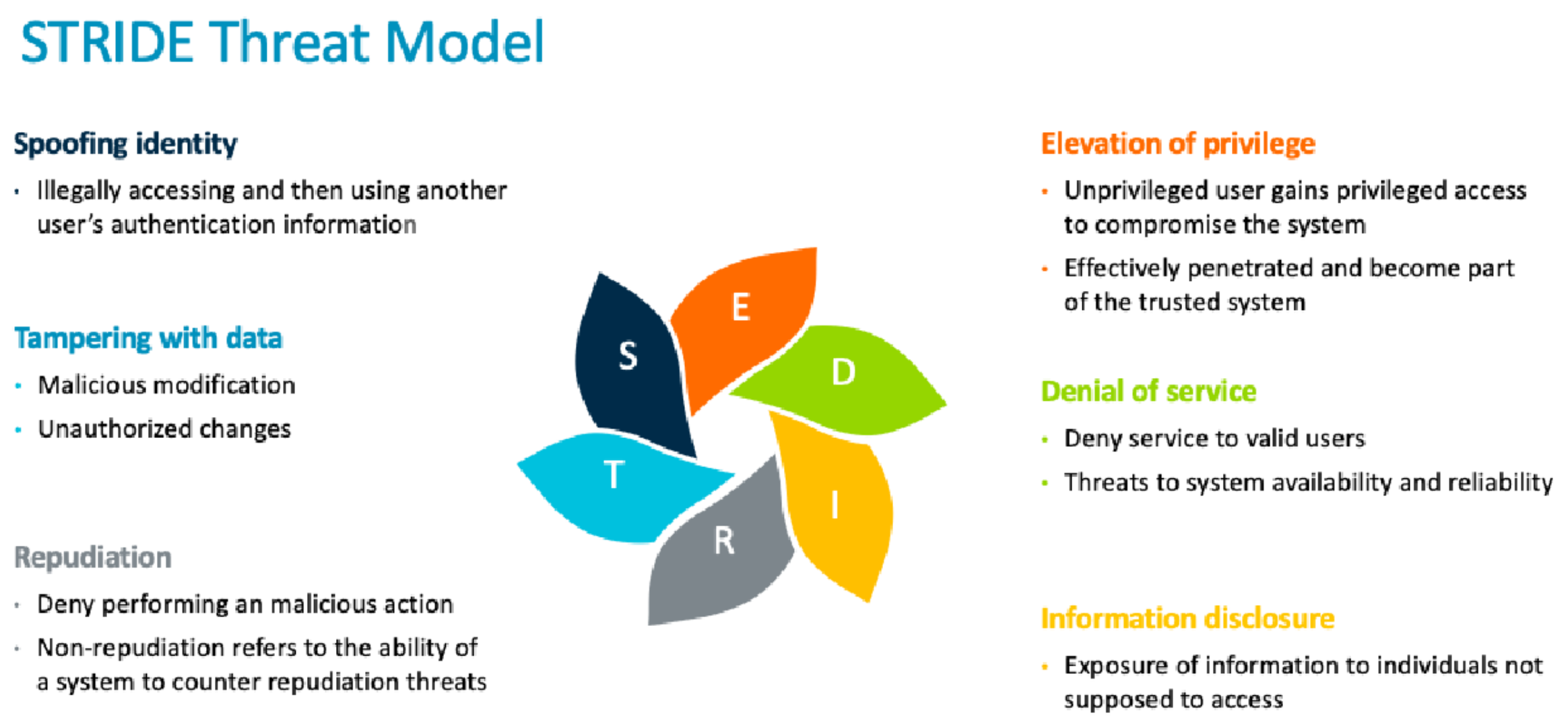

Figure 10: Stride Model [42]

• **FRAP** (facilitated risk analysis process) mostly deals with qualitative aspects. [43]
• **OCTAVE Allegro**: is developed for strategic planning. [44]

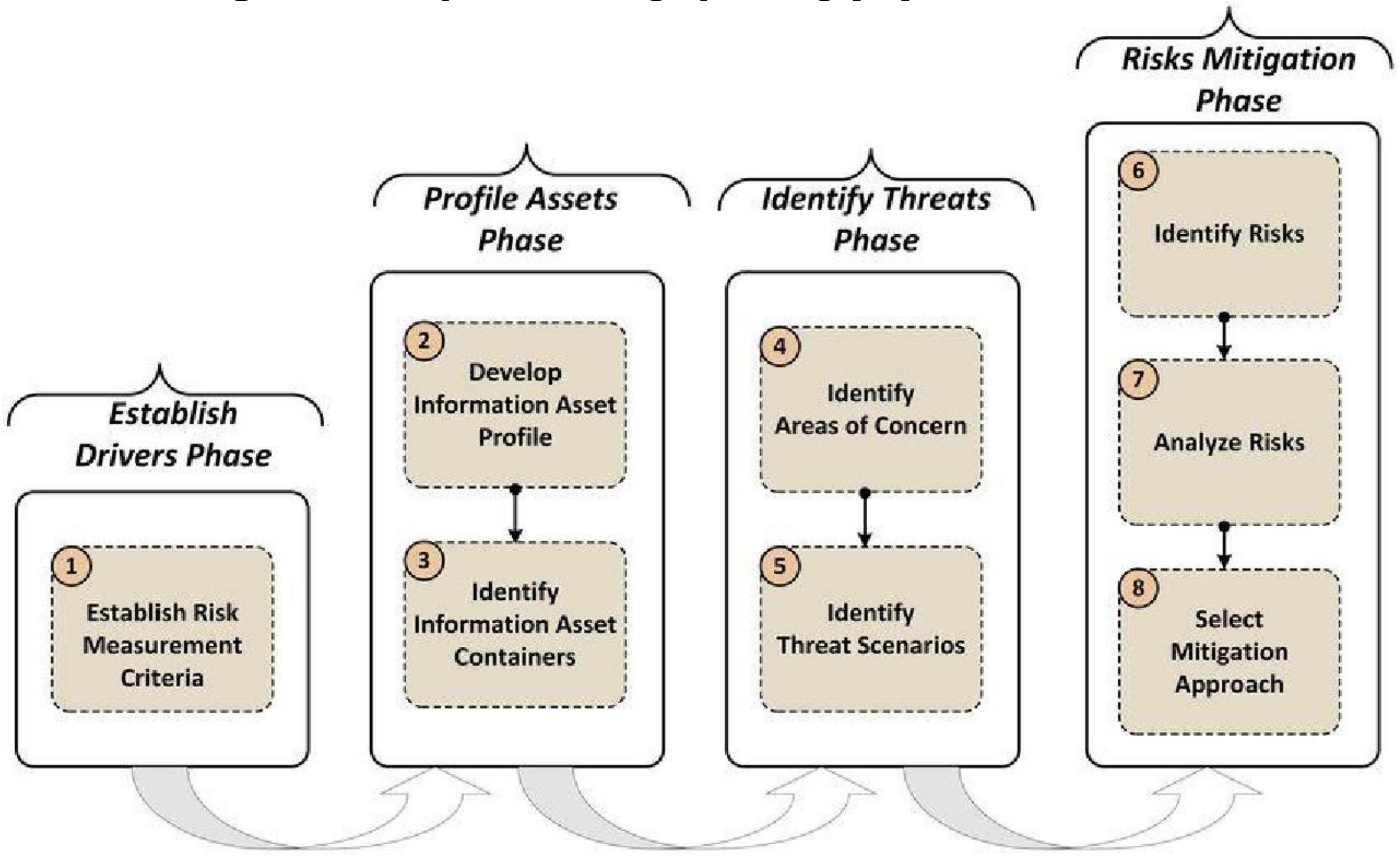

Figure 11: Allergo Octave. [45]
• **Spanning Tree Analysis (SPT)**: It is mainly used to analyze threats. [46]

### 6.2.1 Context Establishment

To establish the context both internal and external contexts are the important parameters to understand.

External context: social, cultural, environmental (like natural calamities and climate issues), political, legal, financial, technological, security, economic factors, external stakeholder's parameters etc.

Internal context: strategic objectives, values, standards, resources available, capacities, business processes, organizational culture, internal stakeholder's factors, etc. Stakeholder's factors mean their perceptions, relationships and expectation to the organization. Our defined system would be evaluated on the basis of following context: Governance, Business continuity & succession planning, Business, Financial, Regulatory, Technology, Human resources, Stakeholder.

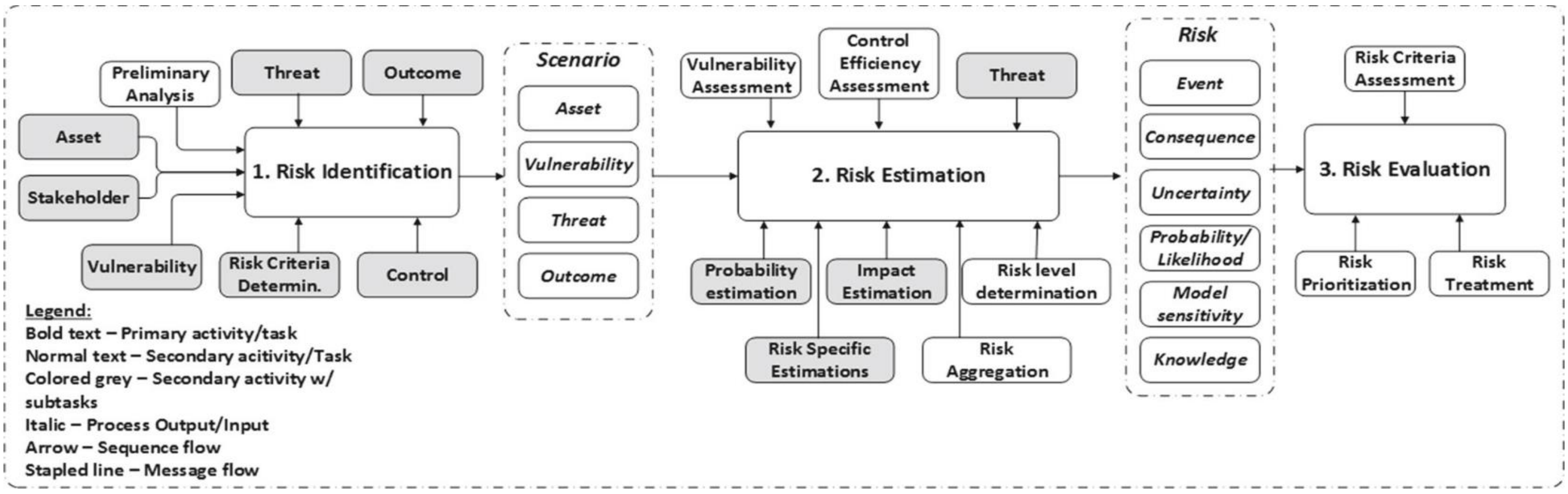

**Figure 12:** The generic output of the risk evaluation is prioritized risks; source [47]

### 6.2.2 Risk Identification

We have to keep in mind that for effective system design, it is urgent to design in such a way where every possible risks would be discussed in order to get out any probable incidents. That's why we must go for comprehensive risk analysis approach which would be started with subsequent phases of context establishment. Now the turn is for risk identification which is sampled in a table here. [48]

### 6.2.3 Risk Analysis

It is associated with risk exposure, likelihood and severity with proper screening by putting adequate attention to get source, history, statistics, W/H info etc. to get the background, vulnerability and credibility.

- **Quantitative methods** is followed by modelling, statistical analysis, decision trees, event tree analysis, probability and consequence analysis.
- **Semi-quantitative methods** focus on a quantitative score to a qualitative approach for instance, high risk, medium risk or low risk.
- **Qualitative methods** deal with brainstorming, specialist knowledge, questionnaires or intuitions.

The relationship among consequences, likelihood, and level of risk is,

Risk = consequence * likelihood. [49]

### 6.2.4 Risk Evaluation

In this step, we have to consider the organizations capacity as it varies from one to another. On top that, it also depends on some factors like: [50]

- Organizational policies
- Own goals and objectives
- Attitude of internal and external stakeholders
- The environment and geographic location
- Managerial structures
- Constraints, budget allocation and resiliency
- Legal limitations
- Approximate maintenance costs
- Possible effect of fines by regulatory authority
- Ethical, humanitarian or moral practices
- Reputation and brand value

### 6.2.5 Risk Treatment

Risk treatment consists of avoiding, reducing, sharing, transfer or accepting. It might differ with risk type, magnitude and other relevant factors.

The following table will give the readers a typical concept of risk assessment for information system where risk management approach has been performed according to ISO 27005. [51]

Table: Typical Risk Assessment Chart

Table 2: Risk Analysis with scoring and ranking

| Serial | Asset | Domain | Threat | Vulnerability | Loss of CIA | Severity 1-5 | Likelihood 1-5 | Risk 1 - 25 | Additional Risk Control Measures | Severity 1-5 | Likelihood 1-5 | Residual 1-25 | Remarks |
|---|---|---|---|---|---|---|---|---|---|---|---|---|---|
| 1 | Web Server | Physical | Being Stolen | Lack of Physical Security | C + A | 5 | 2 | 10 | Improve Physical Security | 5 | 1 | 5 | |
| 2 | Web Server | Physical | Outdated or Non-functional | Lack of Attention or Supervision | A | 2 | 3 | 6 | Improve Supervision | 2 | 1 | 2 | |
| 3 | Web Server | Physical | OS/Server System Fail | Irregular Monitoring on OS | A | 2 | 2 | 4 | Improve Monitoring | 2 | 1 | 2 | |
| 4 | Web Server | Physical | Harddisk Crash | Lack Update of Hardwares | A | 2 | 3 | 6 | More Attention | 2 | 1 | 2 | |
| 5 | Web Server | Virtual | Storage DOS | filling resources | A | 2 | 3 | 6 | Set disk quota's for users | 1 | 1 | 1 | |
| 6 | Web Server | Cloud | Earthquake or War Bombing | Out of Control | A | 1 | 1 | 1 | Selection of Safe Zones | 1 | 1 | 1 | |
| 7 | Access Point | Router/Switch | Attack Surface | DoS | C+I+A | 5 | 3 | 15 | VAPT and Proxy Server using | 5 | 1 | 5 | |
| 8 | Access Point | Physical | Tamper | Raspberry Pi has many open port | C + A + I | 4 | 2 | 8 | Use GPIO with Light Detection to self destruct | 3 | 1 | 3 | |
| 9 | Access Point | Router/Switch | Outdated Firmware | Loosing access control | C + A + I | 5 | 2 | 10 | Regular Firmware Upadating by experts | 5 | 1 | 5 | |
| 10 | Access Point | Virtual | Weak Passwords | unauthorised access | C | 4 | 4 | 16 | Change password, implement policy | 4 | 1 | 4 | |
| 11 | Web Site | Virtual | SQL Injection | Loss of data | C+I+A | 5 | 2 | 10 | Adequate filtering in database | 5 | 1 | 5 | |
| 12 | Website | Virtual | XSS Scripting | Loss of data | C | 5 | 2 | 10 | Securing browser and certificates on token | 5 | 1 | 5 | |
| 13 | Firewall | Virtual | Inside Attack & Security Patches | Loosing access control | C+I+A | 5 | 2 | 10 | Strong administrative and technical control | 5 | 1 | 5 | |
| 14 | DNS | Virtual | Cache Poisoning Attacks | send legitimate requests to malicic | C | 3 | 2 | 6 | Use VPN, HTTPS, Encryption | 3 | 1 | 3 | |
| 15 | Proxy Server | Virtual | SSL-based DDoS attacks | latency and audit isssues | A | 4 | 2 | 8 | Using strong firewall and continuous montitoring | 4 | 1 | 4 | |
| 16 | Workstation Dev | Physical | Physical damage or data leakage | Data loss | C+I+A | 5 | 1 | 5 | Strong physical security system like entry-exit | 5 | 1 | 5 | |
| 17 | User Device | Both | Physical damage or virtual control | loss of data and functionality` | C+I+A | 5 | 1 | 5 | Awareness and Training | 5 | 1 | 5 | |

## 7.0 Risk Management for AV

To improve transportation system in terms of physical security, user friendliness, environmental effects and traffic congestion; the autonomous vehicle (AV) is a delectable addition to science, engineering and technology. The continuous advancement in sensor, computer and tele-communication would take it into scale of performance though it has some new challenges like cyber-security and interaction with other non-autonomous vehicles. The next few topics will make sense to the reader on several risks associated with the behavior of AV in mixed traffic environment considering the cyber-security aspects on the basis of published scientific articles and other relevant data sources. Low-power lasers and pulse generator can build up a system to make AV sensors or lidar fool. Hacker can take controls of brakes, components, other elements remotely. So it is a must to examine all possible risks associated with AV to prevent any future accidents. [52]

The purpose of the risk analysis is to clinch a safer system by protecting it from inherent threats of failure from any existing vulnerabilities. The fault tree analysis is used on identifying the chronological and sequential catastrophe of a system like designing of aircraft processes and nuclear power plant design, and analyzing of risks of AV in Mixed Traffic Streams. The difference between the actual outputs and the expected outputs quantifies the failure. The fault tree analysis method can provide the minimum path for single component or full vehicle failure.

A safe and reliable transportation system can be obtained by comprehensive risk assessment which is the concentration of several researchers for individual subsystem of AV. Surveyed motion prediction and relevant risk assessment methods usually look for collision prediction which is costly in case of improving efficiency to consider physical parameters of AV, weather conditions, road surface conditions speeds approaching an intersection, real trajectories, fatigue level of the driver, data from vehicle sensors lane departure warning and driving assistance subsystem, pedestrian behavior in different traffic scenarios. The Bayesian belief network approach method only consider optical system like camera and lidar.

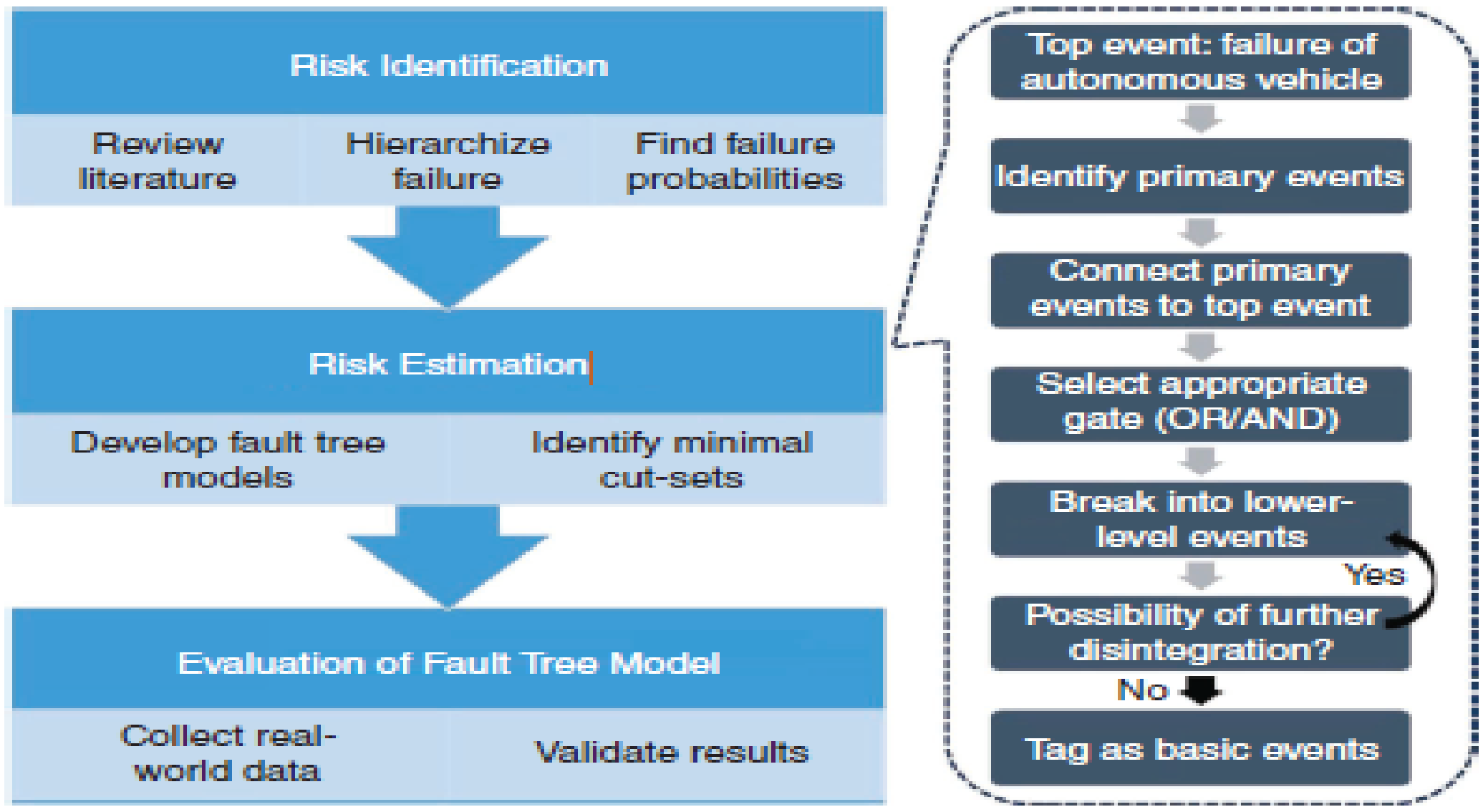

Figure 13: General Research Pattern for Risk Analysis [53]

The above figure can be a sample of approach for risk assessment of an AV which includes risk identification, risk estimation and evaluation of fault tree model.

Table 3: Failure Probabilities, experiment type, methods, and events of Autonomous Vehicular Components [54]

| Basic Event | Description | Methods | Experiment Type | Failure Probability (%) |
|---|---|---|---|---|
| Lidar failure | Laser malfunction, mirror motor malfunction, position encoder failure, overvoltage, short-circuit, optical receiver damages | Bayesian belief network | Simulation | 10.0000 (*48*) |
| Radar failure | Detection curves drawn with respect to signal and noise ratios | Chi-square distribution | Mathematical modeling | 2.0000 (*51*) |
| Camera failure | Foreign particles, shock wave, over-voltage, short-circuit, vibration from rough terrain, etc. | Bayesian belief network | Simulation | 4.9500 (*48*) |
| Software failure | System had to generate outputs from array definition language statements | Extended Markov Bayesian network | Experiment (3,000 runs) | 1.0000 (*52*) |
| Wheel encoder failure | Encoder feedback unable to be transferred, which can cause loss of synchronization of motor stator and rotor positions | Kalman filter | Experiment | 4.0000 (*53*) |
| GPS failure | Real-life tests performed with high-sensitivity GPS in different signal environments (static and dynamic) for more than 14 h | Least squares | Experiment (at 4 locations) | 0.9250 (*54*) |
| Database service failure | Using new empirical approach, connectivity and operability data of a server system were collected. | Generic quorum-system evaluator | Experiment (for 191 days) | 3.8600 (*55*) |
| Communication failure | Wi-Fi: Periodic transmission of 1,000-byte frames (average conditional probability of success after previous success considered) | In IEEE 802.11b network | Experiment (with 10 vehicles) | 5.1250 (*56*) |
| | LTE: Network unavailability during location update in mobility was considered here | Application of CAP theorem | Experiment | 5.8800 (*57*) |
| Integrated platform failure | A two-state model with failure rates was developed to estimate the computer system availability. | Markov chain model | Mathematical modeling | 2.0000 (*58*) |
| Human command error | Three data sets of over 115 months from NASA were analyzed and then validated by three methods (THERP, CREAM, and NARA) to facilitate NASA risk assessment. | Human reliability analysis | Experiment (from December 1998 to June 2008) | 0.0530 (*59*) |
| System failed to detect human command | System unable to detect the accurate acoustic command; driver inputs the wrong command, and system unable to detect wrong commands | Artificial neural networks on clean speech | Experiments (37 subjects: 185 recording) | 1.4000 (*60*) |

The table sated above has magnified all the basic events related to AV along with description. It also demonstrates which method and experiment can be applied for analysis to calculate probabilities of failure. Non-autonomous vehicles crashes usually resulting from reckless driving, tiredness, hardware and distractions which are considered here. Cyclists were responsible for crashes with AV which are also included here. Pedestrian crashes and construction zones are in line with AV for accidents. Weather-related incidents are caused for fog, mist, sleet, snow, dust, rain, severe crosswind, thunderstorm, and smoke. Road conditions like improper lane marking, holes in road, ice formations, fallen tree, and pavement conditions may cause crashes too.

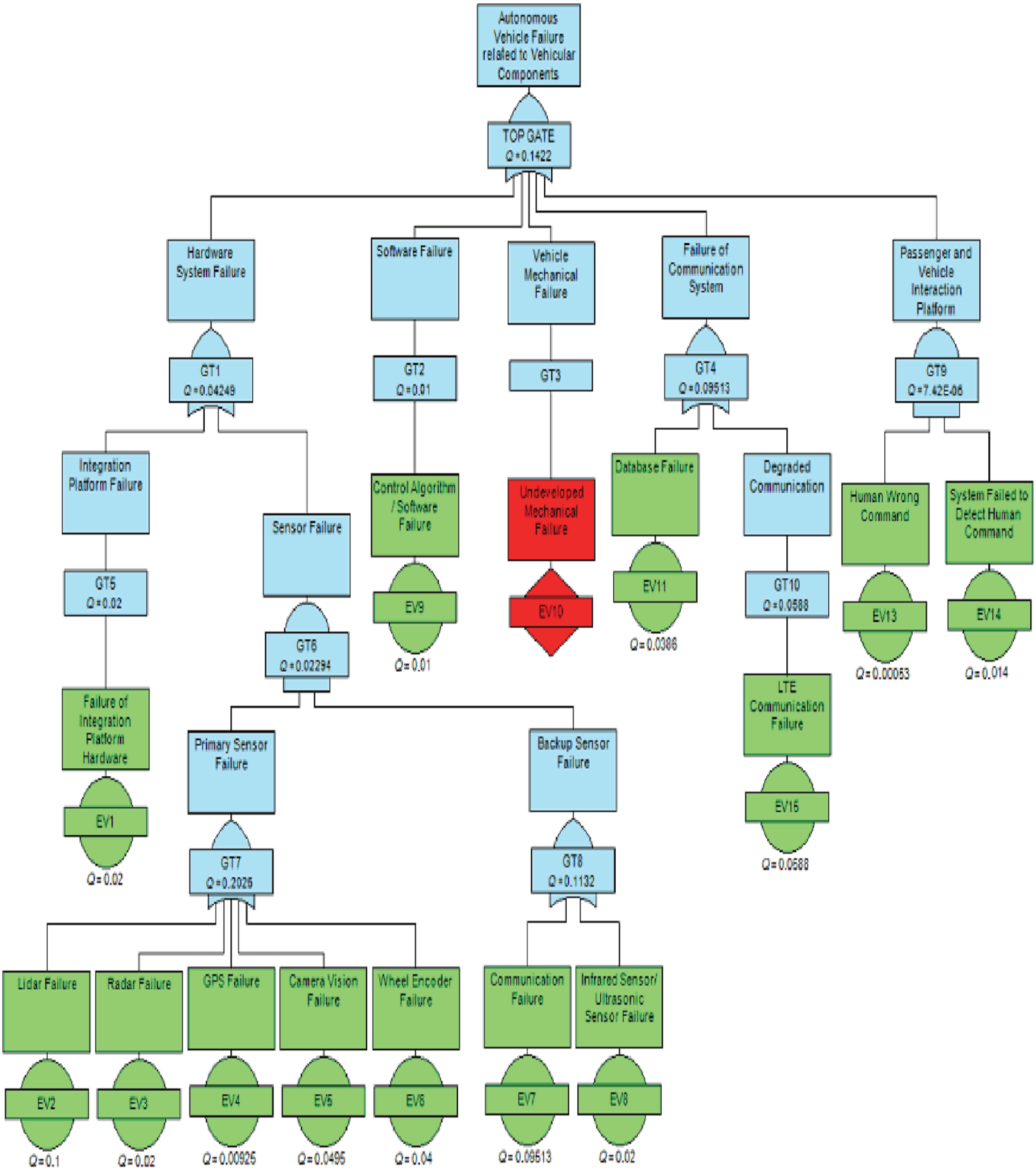

FIGURE 14: Fault tree analysis considering failure related to vehicular components ($Q$ = probability value either inputted into the fault tree or calculated by fault tree analysis). [54]

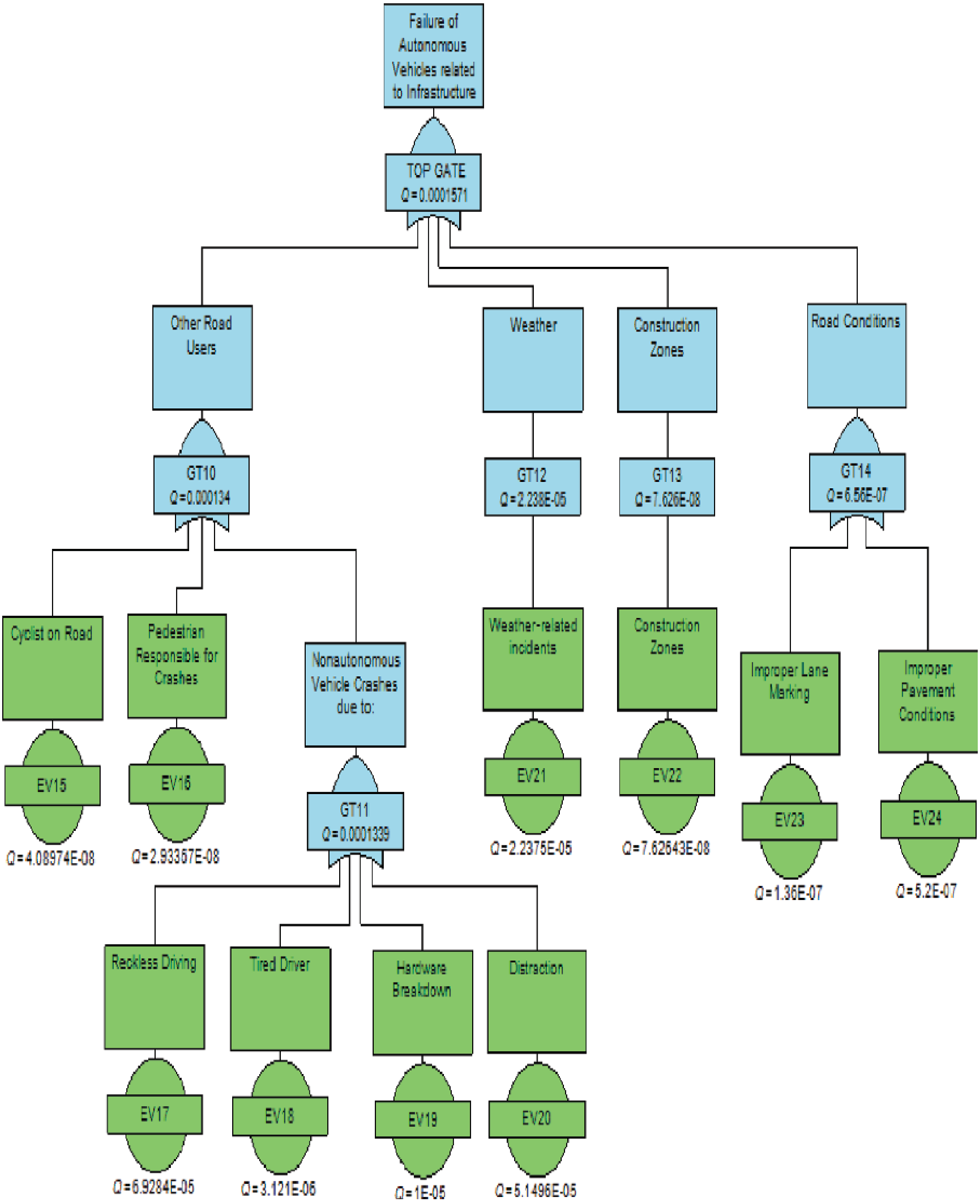

FIGURE 15: Fault tree analysis considering failure related to transportation infrastructure components [55]

A "fixed probability" statistical model can be used to analyze risks on the failure of vehicular components of AV and Isograph "Fault Tree+" software models the distribution of probability for basic event failure. Fault tree for infrastructure component failures for the infrastructure components has been given in figure 3 which is based on fault tree. Combined Fault Tree is obtained by bringing the experimental estimation and simulation modeling together to form a risk hierarchy for upholding risk ranking in terms of score to develop a cut-sets of the main event and it can assist to improve the safety performance of AV.

AV can contribute in a wide range to ensure safe, secure and sustainable transportation system for the next generation where proper risk analysis can be a trigger to reduce loss in real life before any large scale deployment. In depth fault identification by tree analysis might help the developers and researcher to reduce all the probable risks before AV gets implemented in a comprehensive approach through continuous innovations in computing and communication technologies though it might face some limitations in a mixed traffic system for reckless human driver.

Nowadays, researchers are very keen and concentrated to study the cybersecurity of AV where they have categorized threats as high, medium, and low by considering feasibility, probability, frequency, and success rate of every attack. .In order to get anti attack techniques, researchers have widely gone through attack and defense sample of those steps will be discussed in general in upcoming topics.

graphs to realize computer networks' security as a powerful tool in analysis of risks for vulnerable components in AVs. These graphs help to visualize all countermeasures against each risks. [56]

The proposed plain security model can monitor vulnerable components by the means of security nodes. EVITA (E-safety vehicle intrusion protected applications) represents countermeasures in a graph having all security states.

EVITA and CVSS (Common Vulnerability Scoring System) are two widely used methodologies to conduct the risk assessment of AV based on likelihood, impact and severity. Safety, privacy of drivers, operational performance, and financial losses are four important parameters to calculate risks of a vehicle which can be categorized as none, low, medium, and high. [57]

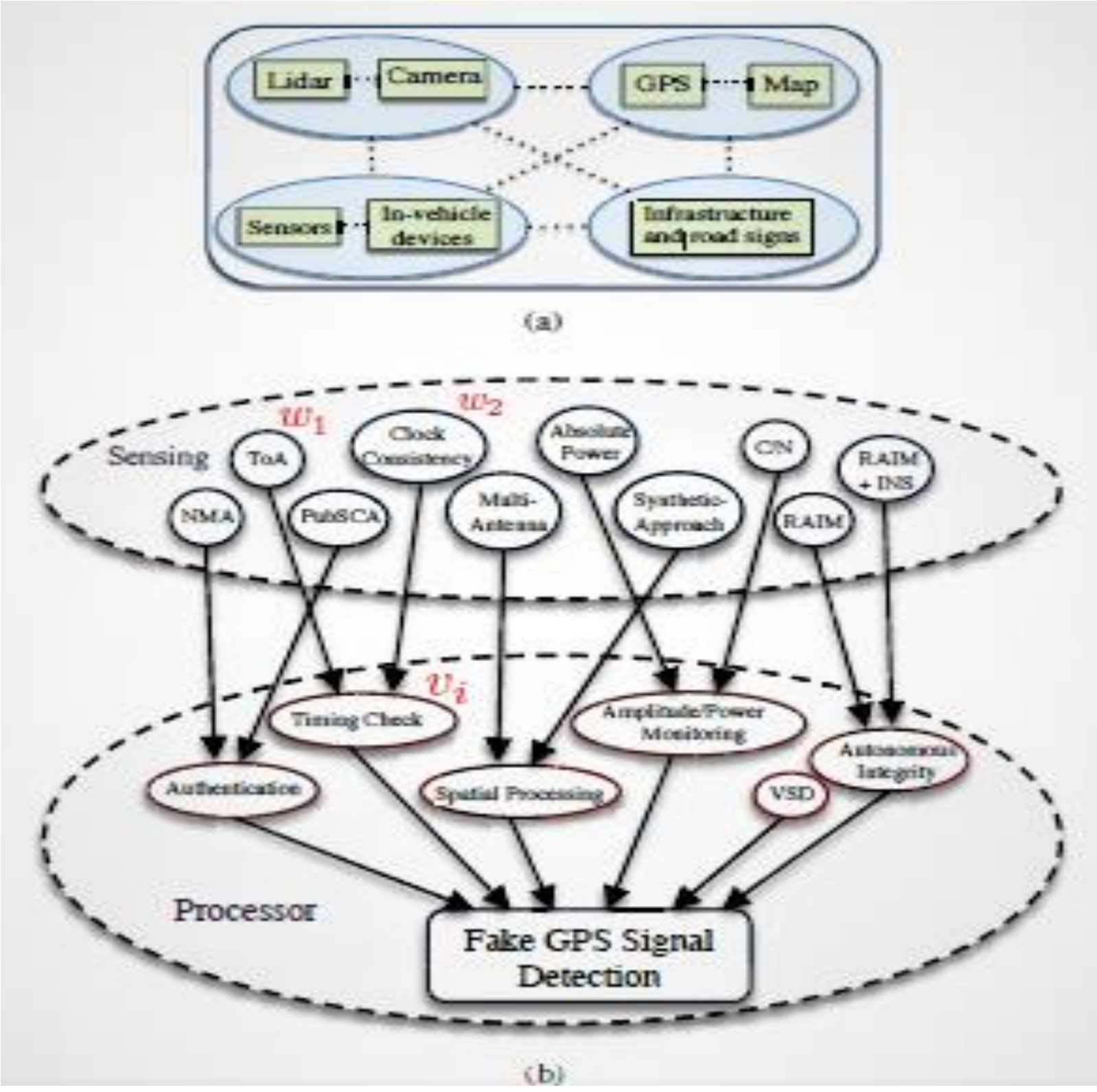

Figure 16: (a) Security monitoring unit, and (b) graphical model for secured GPS component in AV. [58]

Bayesian defense graph has been introduced here to study the cyber related risks and threats of AVs, for example, EVITA has been applied for a framework for GPS signals. However, the likelihood of threats from existing vulnerabilities can be declined to 0:01% by using anti-spoofing techniques which can mobilize future work. [59]

Vehicles Risk Analysis (VeRA) is an appropriate method for the analysis of different risks of single AV and Connected AV (CAV) considering multiple factors (standard SAEJ3061). It is comparatively an easier approach which discussed three important parameters simultaneously and we will also look at a comparison of VeRA with some other models. The knowledge (K) as well as equipment (E) will determine the attack probability (P) has also been represented in the following table.

A formula can be obtained by the combination of attack probability P, severity S and human control H as follows:

Risk value, R= H + S*P

Table 4: VeRA Risk Value calculation using formula.

ATTACK PROBABILITY $P$ CLASSIFICATION, AND CORRESPONDING $K+E$.

| Attack probability $P$ | Description | Corresponding (K+E) |
|---|---|---|
| 1 | Low | $K+E>2$ |
| 2 | Median | $1<K+E\leq 2$ |
| 3 | High | $K+E\leq 1$ |

VERA RISK CLASSIFICATION MATRIX, ACCORDING TO $H$, $S$ AND $P$.

| Human Control $H$ | Severity $S$ | Attack probability $P$ | | |
|---|---|---|---|---|
| | | 1 | 2 | 3 |
| 1 | 1 | 2 | 3 | 4 |
| | 2 | 3 | 5 | 7 |
| | 3 | 4 | 7 | 10 |
| 2 | 1 | 3 | 4 | 5 |
| | 2 | 4 | 6 | 8 |
| | 3 | 5 | 8 | 11 |
| 3 | 1 | 4 | 5 | 6 |
| | 2 | 5 | 7 | 9 |
| | 3 | 6 | 9 | 12 |

Table 5: VeRA Severity Scale

| Severity $S$ | Safety $S_s$ | Privacy $S_p$ | Financial (in €) $S_f$ | Operational $S_o$ |
|---|---|---|---|---|
| 0 | No injuries | No unauthorized access to data | 0<loss<100 | No impact on performance |
| 1 | Light injuries | Access to anonymous data | 100<loss<1000 | Impact not detected by driver |
| 2 | Severe injuries, with survival | Identification of car or driver | 1000<loss<10000 | driver aware of performance degradation |
| 3 | Life threatening, possible death | Driver or car tracking | loss>10000 | Significant impact on performance |

Table 6: Relationship among different models and inter-connections.

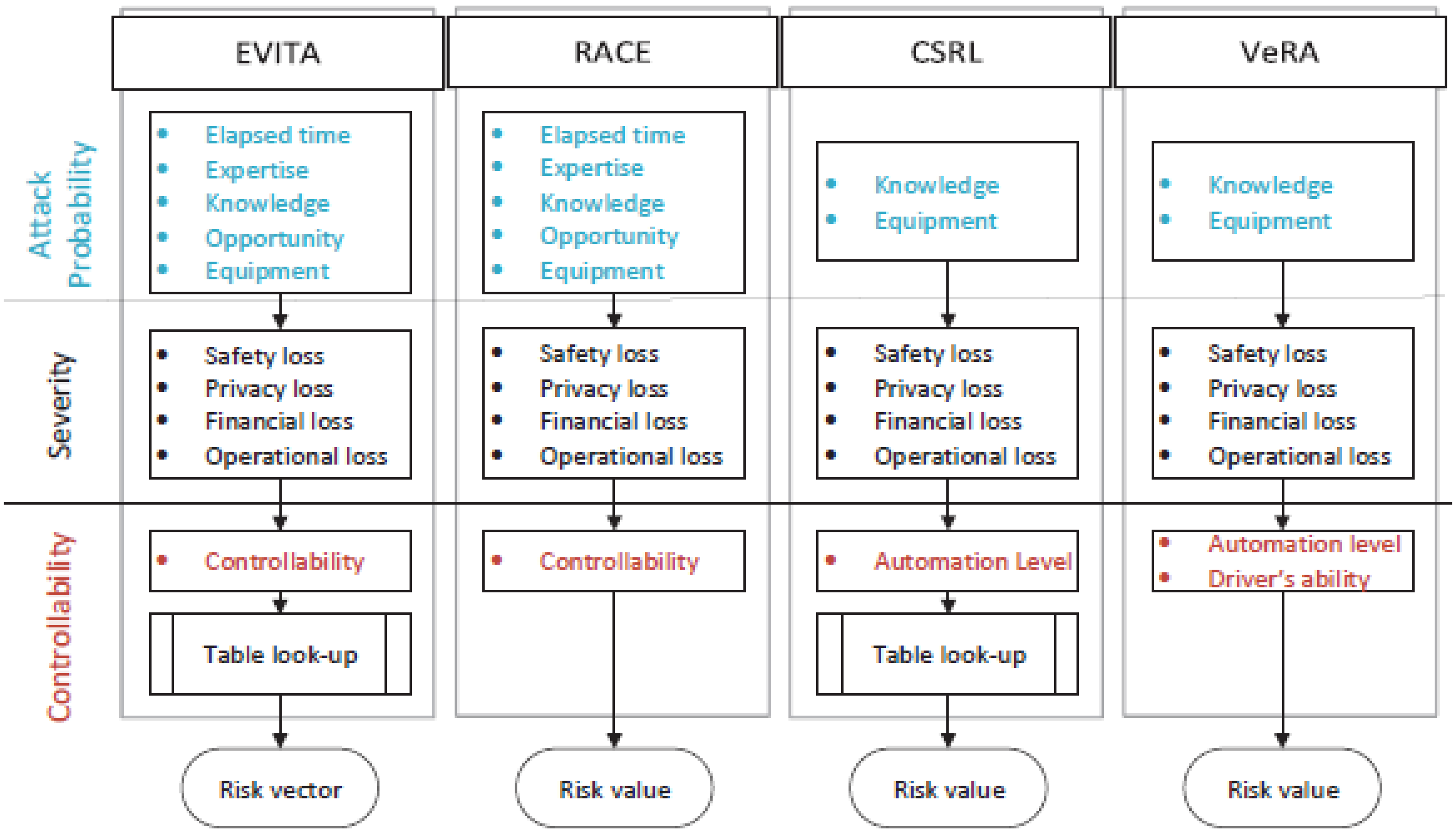

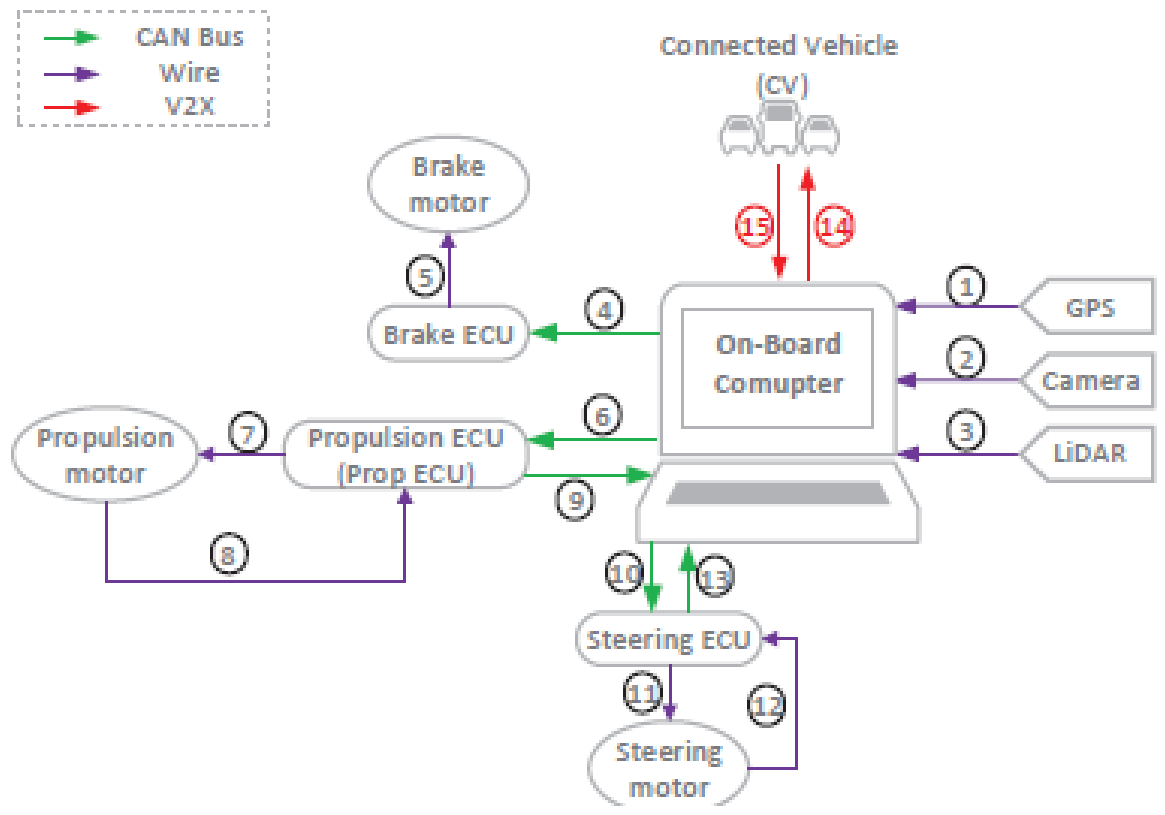

| Function | Required data | Output data |
|---|---|---|
| *Localization* | 3D map / GPS / LiDAR | Pose*, velocity* |
| *Co-Localization* | 3D map / status data (from CV) | CV's location+ |
| *Object Detection and Tracking* | Camera / LiDAR | Recognized object¶ |
| *Co-Object Detection and Tracking* | 3D map / CV's location+ / sensor data (from CV) | Hidden object§ |
| *Mission Generation* | Pose* / CV's location+ / vector map / destination | Global waypoint† |
| *Motion Planning* | Pose* / velocity* / recognized object¶ / hidden object§ / vector map / global waypoint† | Local waypoint‡ |
| *Path Following* | Pose* / velocity* / local waypoint‡ | Target velocity and angle |

The superscripts mark the output data.

Figure 17: Data Acquisition Strategy for VeRA Model

Table 7: Description of different types of attacks.

| ***Detailed attack*** | ***Description*** |
|---|---|
| Alteration | Modify configuration files |
| DoS | Make network link unavailable by flooding the network interface controller with bad traffic packets |
| Eavesdropping | Capture secretly and analyse the exchanged messages |
| Sensor Jamming | Interfere with sensor antennas by saturating their receivers with noise and false formation, e.g., GPS jamming, LiDAR jamming |
| Malware | Run a malware on an equipment (e.g., ECU, external sensor), either by flashing it directly into the ECU memory or using the firmware update process |
| Message injection | Inject self-created information or command on a network link (Ethernet, CAN or wireless) |
| Message suppression | Delete messages on a network link |
| Replay | Replay of past sniffed traffic |
| Sensor spoofing | Manipulate sensor to fake the captured and returned information. This attack includes rogue GPS attack where we generate radio signals to simulate a GPS signal to make the receiver use fake position information (i.e., GPS Spoofing) |
| Sybil | A malicious vehicle transmits the various messages with multiple fake or stolen source identities to other vehicles |
| Falsified entity attack | The attacker passes information to the connected vehicle through a valid network identifier |

RACE (Risk Analysis for Cooperative Engines) is another method of risk analysis specially developed for CAV which is more efficient over EVITA but it requires much effort keeping controllability constant. [60]

Attack probability, severity and vehicle's automation level form US2 method for AVs to determine safety, security and attacks defined by SAE standard J3016 excluding CAVs which extends to the CSRL (Cyber Security Risk Level) for CAVs. The CSRL method is slow and it does not take automation and human control into account. [61]

The international society of automotive engineers (SAE) has found six levels of automation which are demonstrated in the table below. [62]

Table 8: Levels of Autonomy by SAE [63]

| SAE level | Name | Narrative Definition | Execution of Steering and Acceleration/ Deceleration | *Monitoring* of Driving Environment | Fallback Performance of *Dynamic Driving Task* | System Capability *(Driving Modes)* |
|---|---|---|---|---|---|---|
| ***Human driver* monitors the driving environment** | | | | | | |
| 0 | No Automation | the full-time performance by the *human driver* of all aspects of the *dynamic driving task*, even when enhanced by warning or intervention systems | Human driver | Human driver | Human driver | n/a |
| 1 | Driver Assistance | the *driving mode*-specific execution by a driver assistance system of either steering or acceleration/deceleration using information about the driving environment and with the expectation that the *human driver* perform all remaining aspects of the *dynamic driving task* | Human driver and system | Human driver | Human driver | Some driving modes |
| 2 | Partial Automation | the *driving mode*-specific execution by one or more driver assistance systems of both steering and acceleration/ deceleration using information about the driving environment and with the expectation that the *human driver* perform all remaining aspects of the *dynamic driving task* | **System** | Human driver | Human driver | Some driving modes |
| ***Automated driving system* ("system") monitors the driving environment** | | | | | | |
| 3 | Conditional Automation | the *driving mode*-specific performance by an *automated driving system* of all aspects of the dynamic driving task with the expectation that the *human driver* will respond appropriately to a *request to intervene* | System | **System** | Human driver | Some driving modes |
| 4 | High Automation | the *driving mode*-specific performance by an automated driving system of all aspects of the *dynamic driving task*, even if a *human driver* does not respond appropriately to a *request to intervene* | System | System | **System** | Some driving modes |
| 5 | Full Automation | the full-time performance by an *automated driving system* of all aspects of the *dynamic driving task* under all roadway and environmental conditions that can be managed by a *human driver* | System | System | System | **All driving modes** |

ISO 26262 process is currently widely used for risk analysis of AV which consists of

- Item Definition,
- Hazard and Risk Analysis,
- ASIL Assignment
- Determination of safety goals. [64]

Table 9: Application of ISO 26262 with STPA for Engine Systems. [65] [66]

| Item | Malfunction | Driving Situation | Risk Scenario | Hazard |
|---|---|---|---|---|
| Engine System (S7) | f1: Accelerates more than intended | Urban driving environments with medium speeds between 25 mph and 40 mph | Traffic en-route or presence of obstacles | Collision possible with infrastructure or pedestrians |
| | f2: Missing feedback from sensors regarding vehicle state | Freeway at high speeds (> 40 mph) | Heavy obstacle present at $< 300m$ | Collision possible with traffic or infrastructure |
| | f3: Erroneous throttle valve signal | Freeway at high speeds (> 40 mph) | Minimum preferred distance to obstacle exceeded | Collision possible with traffic or infrastructure |

| Safety Goal |
|---|
| SG1: Avoid unintended increase in engine torque in Urban areas beyond t1 time period |
| SG2: Avoid unintended decrease in engine torque in Urban areas beyond t2 time period |
| SG3: Avoid unintended acceleration in urban areas beyond t1 time period |
| SG4: Avoid unintended deceleration beyond t2 time period |
| SG5: Ensure integrity in sensor feedback |

## 8. Issues:

The Internet of Autonomous Vehicles (IoAV)IoAV platform upholds improved transportation system besides of better fuel economy, latest safety standards, minimum road accidents, and more comfortable journey experience. However, it is associated with multiple issues from device design, data management, social, ethical, professional, legal, security issues.

### 8.1 Ethical Issues

The ethical concepts of HAVs [highly autonomous vehicles] are developed and deployed by manufacturers, consumers, government, and stakeholders which is a critical concept having so many dilemmas. It can only be trained by policy imposing but still contains conflict among:

- an individual's interest
- the community's interest
- protecting people from harm
- providing people healthy lives.

To have a specific example of such dilemma, the readers are requested to the trolley problem. [67]

The common ethical stand for all information system is to provide safety and reliability in all types information and data handling though it varies from system to system like open or close networks, LANs, ISPs, backbone sites etc.

Our system administrator would definitely focus on all possible threats to provide privacy, liability, patent and copyright laws. It has to ensure confidentiality, trade secrets, fraud and misuse cases, decision making capacity based on ethics and responsibility. [68]

### 8.2 Environmental Issues:

The environmental issues which are related to mainly with energy consumption, car emissions, expired parts reclining etc. So the engineer, designer and researcher of AV must consider to ensure low rate of fuel consumption, enhanced use of renewable energy, invent new technology for less carbon emissions, high quality road design to lessen frictional loss, finding shortest path of travelling.

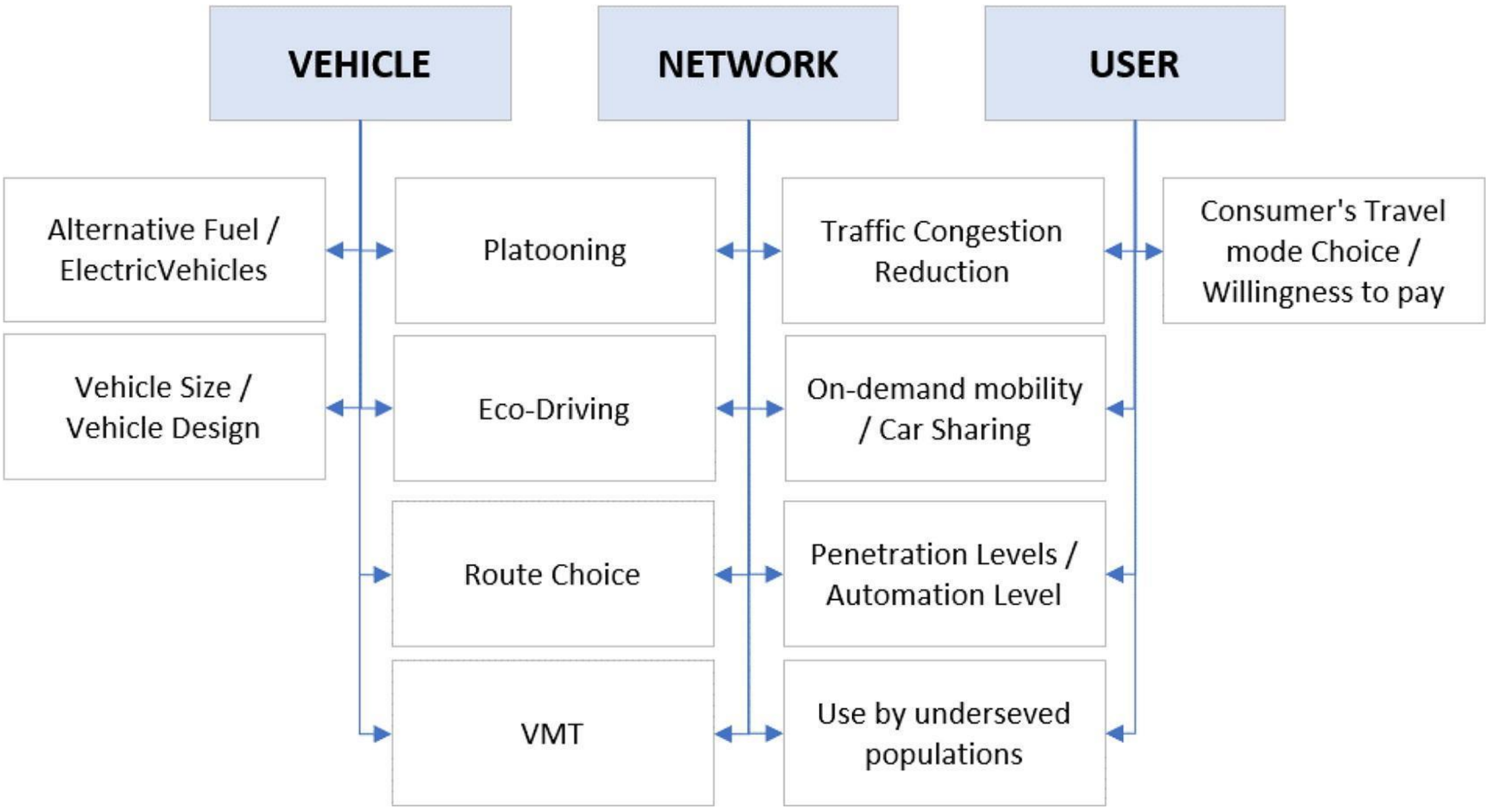

Figure 18: Key elements for environmental concern [69]

### 8.3 Legal Issues

It varies with locality, state, region, country, policy and government. However, AV must deal with two different aspects of law, one is road transport law and another information system domain law. Our defined system is related to road, transports, information, information system, data, finance, cyber security, computer system, people etc. So we can term it as multi-disciplinary sector mostly deals with public data. So several legal frameworks are available here based socio-economic and geo-political circumstances like GDPR, DPA, BSC, IEEE, BCS, and IET etc. We have to abide by all those issues to operate our system as a white one. [70] [71] [72]

- Traffic Acts
- Vehicle Acts
- General Data Protect Rule (GDPR)
- Data Protection Act
- Road transport law

### 8.4 Professional Issues

Any professional who are related to AV must maintain the moral and ethical values along with the existing law and regulation of both data security, automobile design and road system. It includes all persons like driver, robots, companies, engineer, data administrator etc. who are somehow dealing AV operations. Any incidents must spread liabilities to all of those personnel to ensure accountability. [73]

### 8.5. Social Issues

One of the social dilemmas that occurred with AV systems is the question, if 10 pedestrians appeared on the vehicles path and the vehicle is not able to break early enough, should the algorithm in the autonomous vehicle save the 10 pedestrians or the passenger? One study by Bonnefon et al., 2016 investigated human morality when it comes to those two algorithms, the utilitarian approach that chooses to save a greater number of lives, and the non-utilitarian approach that chooses to save the passenger (passenger-protective). Given this dilemma, participants were required to rate which action they thought was more moral. The result of this survey was that the participants found the utilitarian algorithm as the moral choice. On the other hand, when another survey was taken where the participants were asked about the likelihood of purchasing the vehicle of each algorithm, the study found that people rated the vehicle with the non-utilitarian algorithm more likely to purchase. A social dilemma arises due to these surprising results. There is a temptation that occurs to act according to self-interest, which usually leads to the worst results for everyone involved [29].

The research that was done created uncertainty in the number of participants. It is possible that the responses were different for each question because of limited access to moral utilitarian information. One example would be that the participant envisioned themselves as the driver, however even the driver would also at some point be the pedestrian. To eliminate the conflict in the responses, full description of the utilitarian approach should be given, including the utilitarian tasks and their consequences.

There is a psychological uncertainty found in the different utilitarian choices for scenarios that are morally sensitive. The uncertainty can be reduced by comprehensive moral questions and tasks. When it came to Bonnefon's study, the utilitarian information was limited. In the study participants were asked to envision themselves as the passenger, therefore the participant given one side of the situation. A proposal was made that implies that asking the participants to envision themselves as the pedestrian too would help with solving the issue of uncertainty. This way the participant can have both views of the situation, the passengers view and the pedestrian view [30].

## 9. Conclusion

The fourth Industrial revolution demands the automation where the autonomous vehicles is a key element. The prime aspects has discussed in this chapter shortly. Smart transportation does not literally mean AV but it is the synonym of AV in depth realization. The urban planner will work to build smarter cities keeping the idea of AV in mind. On the other hand, there are lots of challenges for the maintenance of AV in terms of road designing, fuel managing, security handling etc. The security concerns are discussed giving emphasize and later on, the issues are raised. So it is our belief that, the readers will get passion on AV while going through this chapter. As it is a chapter merely so it might have limitations. Because, only AV can be covered in several books. However, as a starter it would be a good piece to have insights on AV.

**References**:

[1] Schwarz, Chris, Geb Thomas, Kory Nelson, Michael McCrary, Nick Sclarmann, and Matthew Powell. Towards autonomous vehicles. No. MATC-UI: 117. Mid-America Transportation Center, 2013.

[2] Fagnant, Daniel J., and Kara Kockelman. "Preparing a nation for autonomous vehicles: opportunities, barriers and policy recommendations." *Transportation Research Part A: Policy and Practice* 77 (2015): 167-181.

[3] Schwarting, Wilko, Javier Alonso-Mora, and Daniela Rus. "Planning and decision-making for autonomous vehicles." *Annual Review of Control, Robotics, and Autonomous Systems* (2018).

[4] S. Kato, E. Takeuchi, Y. Ishiguro, Y. Ninomiya, K. Takeda and T. Hamada, "An Open Approach to Autonomous Vehicles," in IEEE Micro, vol. 35, no. 6, pp. 60-68, Nov.-Dec. 2015, doi: 10.1109/MM.2015.133.

[5] J. Kocić, N. Jovičić and V. Drndarević, "Sensors and Sensor Fusion in Autonomous Vehicles," 2018 26th Telecommunications Forum (TELFOR), Belgrade, 2018, pp. 420-425, doi: 10.1109/TELFOR.2018.8612054.

[6] Zehang Sun, G. Bebis and R. Miller, "On-road vehicle detection: a review," in IEEE Transactions on Pattern Analysis and Machine Intelligence, vol. 28, no. 5, pp. 694-711, May 2006, doi: 10.1109/TPAMI.2006.104.

[7] S. M. Patole, M. Torlak, D. Wang and M. Ali, "Automotive radars: A review of signal processing techniques," in IEEE Signal Processing Magazine, vol. 34, no. 2, pp. 22-35, March 2017, doi: 10.1109/MSP.2016.2628914.

[8] Khader M. and Cherian D., "An Introduction to Automotive LIDAR," Texas Instruments, May. 2020. Accessed on: Dec. 27, 2020. [Online]. Available: https://www.ti.com/lit/wp/slyy150a/slyy150a.pdf?ts=1609169652732&ref_url=https%253A%252F%252Fwww.google.com%252F

[9] R. Thakur, "Scanning LIDAR in Advanced Driver Assistance Systems and Beyond: Building a road map for next-generation LIDAR technology," in IEEE Consumer Electronics Magazine, vol. 5, no. 3, pp. 48-54, July 2016, doi: 10.1109/MCE.2016.2556878.

[10] S. Kato, E. Takeuchi, Y. Ishiguro, Y. Ninomiya, K. Takeda and T. Hamada, "An Open Approach to Autonomous Vehicles," in IEEE Micro, vol. 35, no. 6, pp. 60-68, Nov.-Dec. 2015, doi: 10.1109/MM.2015.133.

[11] J. Levinson et al., "Towards fully autonomous driving: Systems and algorithms," 2011 IEEE Intelligent Vehicles Symposium (IV), Baden-Baden, 2011, pp. 163-168, doi: 10.1109/IVS.2011.5940562.

[12] Durrant-Whyte, Hugh, David Rye, and Eduardo Nebot. "Localization of autonomous guided vehicles." *Robotics Research*. Springer, London, 1996. 613-625.

[13] J. Levinson, M. Montemerlo and S. Thrun, "Map-Based Precision Vehicle Localization in Urban Environments", *Robotics Science and Systems*, 2007.

[14] A. Teichman and S. Thrun, "Practical object recognition in autonomous driving and beyond," Advanced Robotics and its Social Impacts, Half-Moon Bay, CA, 2011, pp. 35-38, doi: 10.1109/ARSO.2011.6301978.

[15] A. Teichman, J. Levinson and S. Thrun, "Towards 3D object recognition via classification of arbitrary object tracks," 2011 IEEE International Conference on Robotics and Automation, Shanghai, 2011, pp. 4034-4041, doi: 10.1109/ICRA.2011.5979636.

[16] Goldberg, Ken, and Ben Kehoe. "Cloud robotics and automation: A survey of related work." *EECS Department, University of California, Berkeley, Tech. Rep. UCB/EECS-2013-5* (2013).

[17] D. Hunziker, M. Gajamohan, M. Waibel and R. D'Andrea, "Rapyuta: The roboearth cloud engine", *Robotics and Automation (ICRA) 2013 IEEE International Conference*, pp. 438-444, 2013.

[18] J. Lee, J. Wang, D. Crandall, S. Šabanović and G. Fox, "Real-Time, Cloud-Based Object Detection for Unmanned Aerial Vehicles," 2017 First IEEE International Conference on Robotic Computing (IRC), Taichung, 2017, pp. 36-43, doi: 10.1109/IRC.2017.77.

[19] B. V. Philip, T. Alpcan, J. Jin and M. Palaniswami, "Distributed Real-Time IoT for Autonomous Vehicles," in IEEE Transactions on Industrial Informatics, vol. 15, no. 2, pp. 1131-1140, Feb. 2019, doi: 10.1109/TII.2018.2877217.

[20] D. Minovski, C. Åhlund and K. Mitra, "Modeling Quality of IoT Experience in Autonomous Vehicles," in IEEE Internet of Things Journal, vol. 7, no. 5, pp. 3833-3849, May 2020, doi: 10.1109/JIOT.2020.2975418.

[21] J. Dilley, A. Poelstra, J. Wilkins, M. Piekarska, B. Gorlick and M. Friedenbach, "Strong federations: An interoperable blockchain solution to centralized third party risks", *arXiv preprint arXiv:1612.05491*, 2016.

[22] H. Guo, E. Meamari and C. Shen, "Blockchain-inspired Event Recording System for Autonomous Vehicles," 2018 1st IEEE International Conference on Hot Information-Centric Networking (HotICN), Shenzhen, 2018, pp. 218-222, doi: 10.1109/HOTICN.2018.8606016.

[23] *Umsl.edu*, 2020. [Online]. Available: http://www.umsl.edu/~siegelj/information_theory/projects/HashingFunctionsInCryptography.html. [Accessed: 30- Dec- 2020].

[24] J. B. Kenney, "Dedicated Short-Range Communications (DSRC) Standards in the United States," in Proceedings of the IEEE, vol. 99, no. 7, pp. 1162-1182, July 2011, doi: 10.1109/JPROC.2011.2132790.

[25] G. He and S. Ma, "A study on the short-term prediction of traffic volume based on wavelet analysis", *Proc. IEEE Int. Conf. Intell. Transp. Syst.*, pp. 731-735, 2002.

[26] D. Chen, "Research on Traffic Flow Prediction in the Big Data Environment Based on the Improved RBF Neural Network," in IEEE Transactions on Industrial Informatics, vol. 13, no. 4, pp. 2000-2008, Aug. 2017, doi: 10.1109/TII.2017.2682855.

[27] Karaboga, Dervis, and Bahriye Basturk. "On the performance of artificial bee colony (ABC) algorithm." *Applied soft computing* 8, no. 1 (2008): 687-697.

[28] S. Okdem, D. Karaboga and C. Ozturk, "An application of Wireless Sensor Network routing based on Artificial Bee Colony Algorithm," 2011 IEEE Congress of Evolutionary Computation (CEC), New Orleans, LA, 2011, pp. 326-330, doi: 10.1109/CEC.2011.5949636.

[29] Bonnefon, J.-F., Shariff, A., and Rahwan, I. (2016). The social dilemma of autonomous vehicles. *Science* 352, 1573–1576. doi: 10.1126/science.aaf2654

[30] Martin, Rose, Ivaylo Kusev, Alex J. Cooke, Victoria Baranova, Paul Van Schaik, and Petko Kusev. "Commentary: The social dilemma of autonomous vehicles." *Frontiers in Psychology* 8 (2017): 808.

[31] www.analyticsinsight.net/top-10-autonomous-vehicle-companies-watch-2020/

[32] www.alliedmarketresearch.com/autonomous-vehicle-market

[33] builtin.com/transportation-tech/self-driving-car-companies

[34]Patiño, S., Solís, E.F., Yoo, S.G. and Arroyo, R., 2018, April. ICT risk management methodology proposal for governmental entities based on ISO/IEC 27005. In *2018 International Conference on eDemocracy & eGovernment (ICEDEG)* (pp. 75-82). IEEE.

[35] Wangen, G., Hallstensen, C. and Snekkenes, E., 2018. A framework for estimating information security risk assessment method completeness. *International Journal of Information Security*, *17*(6), pp.681-699.

[36] Information technology –security techniques –information security risk management. Iso, 2011

[37] V. Agrawal, "A Framework for the Information Classification in ISO 27005 Standard," 2017 IEEE 4th International Conference on Cyber Security and Cloud

[38] Lepofsky, R. (2014) *the manager's guide to web application security: a concise guide to the weaker side of the web*. Berkeley, CA: Apress (The expert's voice in security). doi: 10.1007/978-1-4842-0148-0.

[39] Han, Z. *et al.* (2016) “Risk Assessment of Digital Library Information Security: A Case Study,” *The Electronic Library*, 34(3), pp. 471–487. doi: 10.1108/EL-09-2014-0158.

[40] El Fray, I., 2012, September. A comparative study of risk assessment methods, MEHARI & CRAMM with a new formal model of risk assessment (FoMRA) in information systems. In IFIP International Conference on Computer Information Systems and Industrial Management (pp. 428-442). Springer, Berlin, Heidelberg.

[41] Gregory, R. and Mendelsohn, R., 1993. Perceived risk, dread, and benefits. *Risk Analysis*, *13*(3), pp.259-264.

[42] Jiang, L., Chen, H. and Deng, F., 2010, May. A security evaluation method based on STRIDE model for web service. In 2010 2nd International Workshop on Intelligent Systems and Applications (pp. 1-5). IEEE.

[43] Peltier, T.R., 2000. Facilitated risk analysis process (FRAP). Auerbach Publication, CRC Press LLC.

[44] Ali, B. and Awad, A.I., 2018. Cyber and physical security vulnerability assessment for IoT-based smart homes. sensors, 18(3), p.817.

[45] Keating, C.G., 2014. Validating the octave allegro information systems risk assessment methodology: a case study. 28

[46] Huang, G., Li, X. and He, J., 2006, May. Dynamic minimal spanning tree routing protocol for large wireless sensor networks. In *2006 1ST IEEE Conference on Industrial Electronics and Applications* (pp. 1-5). IEEE.

[47] Alghamdi, B.S., Elnamaky, M., Arafah, M.A., Alsabaan, M. and Bakry, S.H., 2019. A Context Establishment Framework for Cloud Computing Information Security Risk Management Based on the STOPE View. *IJ Network Security*, *21*(1), pp.166-176.

[48] Chou, D.C., 2013. Risk identification in Green IT practice. *Computer Standards & Interfaces*, *35*(2), pp.231-237.

[49] Aven, T., 2015. *Risk analysis*. John Wiley & Sons.

[50] Masuoka, Y., Naono, K. and Kameyama, S., Hitachi Ltd, 2004. *Network monitoring method for information system, operational risk evaluation method, service business performing method, and insurance business managing method*. U.S. Patent Application 10/629,920.

[51] Sheehan, B., Murphy, F., Ryan, C., Mullins, M. and Liu, H.Y., 2017. Semi-autonomous vehicle motor insurance: A Bayesian Network risk transfer approach. *Transportation Research Part C: Emerging Technologies*, *82*, pp.124-137.

[52] Bhavsar, P., Das, P., Paugh, M., Dey, K. and Chowdhury, M., 2017. Risk analysis of autonomous vehicles in mixed traffic streams. *Transportation Research Record*, *2625*(1), pp.51-61.

[53] Bhavsar, P., Das, P., Paugh, M., Dey, K. and Chowdhury, M., 2017. Risk analysis of autonomous vehicles in mixed traffic streams. *Transportation Research Record*, *2625*(1), pp.51-61.

[54] Ullman, G. L., M. D. Finley, J. E. Bryden, R. Srinivasan, and F. M. Council. NCHRP Report 627: Traffic Safety Evaluation of Nighttime and Daytime Work Zones. Transportation Research Board, Washington, D.C., 2008

[55] Dezfuli, H., A. Benjamin, C. Everett, G. Maggio, M. Stamatelatos, and R. Youngblood. NASA Risk Management Handbook. Publication NASA/SP-2011-3422. NASA, 2011.

[56] A. Roy, D. S. Kim, and K. S. Trivedi, “Attack countermeasure trees (act):towards unifying the constructs of attack and defense trees,” Security and Communication Networks, vol. 5, no. 8, pp. 929–943, 2012.

[57] O. Henniger, L. Apvrille, A. Fuchs, Y. Roudier, A. Ruddle, and B. Weyl,“Security requirements for automotive on-board networks,” in 9th InternationalConference on Intelligent Transport Systems Telecommunications,(ITST), 2009, pp. 641–646.

[58J. Petit and S. E. Shladover, “Potential cyberattacks on automated vehicles,” IEEE Transactions on Intelligent Transportation Systems ,vol. 16, no. 2, pp. 546–556, April 2015.

[59] Behfarnia, A. and Eslami, A., 2019, September. Local voting games for misbehavior detection in VANETs in presence of uncertainty. In *2019 57th Annual Allerton Conference on Communication, Control, and Computing (Allerton)* (pp. 480-486). IEEE.

[60] A. Boudguiga, A. Boulanger, P. Chiron, W. Klaudel, H. Labiod, andJ.-C. Seguy, “Race: Risk analysis for cooperative engines,” in 7thInternational Conference on New Technologies, Mobility and Security(NTMS). Paris, France: IEEE, July 2015.

[61] Society of Automotive Engineers (SAE), SAE-J3016: Taxonomy and Definitions for terms Related to Driving Automation Systems for On-Road Motor Vehicles, Sep 2016.

[62] G. Sabaliauskaite, J. Cui, L. S. Liew, and F. Zhou, “Integrated safety and cybersecurity risk analysis of cooperative intelligent transport systems,” in SCIS-ISIS, Toyama, Japan, pp. 723–728, Dec 2018.

[63] Jeon, S.H., Cho, J.H., Jung, Y., Park, S. and Han, T.M., 2011, February. Automotive hardware development according to ISO 26262. In *13th International Conference on Advanced Communication Technology (ICACT2011)* (pp. 588-592). IEEE.

[64] The traditional ISO 26262 can be applied along with Hazard and Risk Analysis (HARA), System Theoretic Process and Analysis (STPA) and Automotive Safety Integrity Level (ASIL) for every components of AV separately, as for example for engine systems.

[65] Shastry, A.K., 2018. *Functional safety assessment in autonomous vehicles* (Doctoral dissertation, Virginia Tech).

.

[66] A. Nanda, D. Puthal, J. J. P. C. Rodrigues and S. A. Kozlov, "Internet of Autonomous Vehicles Communications Security: Overview, Issues, and Directions," in IEEE Wireless Communications, vol. 26, no. 4, pp. 60-65, August 2019, doi: 10.1109/MWC.2019.1800503.

[67] Fleetwood, J., 2017. Public health, ethics, and autonomous vehicles. *American journal of public health*, *107*(4), pp.532-537.

[68] Duncan Langford (1997) “Ethical Issues in Network System Design,” *Australasian Journal of Information Systems*, 4. doi: 10.3127/ajis.v4i2.367.

[69] Kopelias, P., Demiridi, E., Vogiatzis, K., Skabardonis, A. and Zafiropoulou, V., 2020. Connected & autonomous vehicles–Environmental impacts–A review. *Science of the total environment*, *712*, p.135237.

[70] Takeyoshi Imai (2019) “Legal Regulation of Autonomous Driving Technology: Current Conditions and Issues in Japan,” *IATSS Research*, 43(4), pp. 263–267. doi: 10.1016/j.iatssr.2019.11.009.

[71] Maurer, M. *et al.* (eds) (2016) *Autonomous driving : technical, legal and social aspects*. Berlin: Springer OPen. doi: 10.1007/978-3-662-48847-8.

[72] Grama, J. L. (2011) *Legal issues in information security*. Sudbury, Mass.: Jones & Bartlett Learning (Jones & Bartlett Learning information systems security & assurance series). Available at: INSERT-MISSING-URL (Accessed: December 6, 2020).

[73] Liu, N., Nikitas, A. and Parkinson, S., 2020. Exploring expert perceptions about the cyber security and privacy of Connected and Autonomous Vehicles: A thematic analysis approach. *Transportation Research Part F: Traffic Psychology and Behaviour*, *75*, pp.66-86.